\documentclass[a4paper,11pt]{article}
\pdfoutput=1 

\usepackage{jcappub,wrapfig,verbatim,amsthm} 
                     
\usepackage[utf8]{inputenc}

\newcommand{\Vg}{V^\mu{}_\nu}
\newcommand{\Vf}{\widetilde{V}^\mu{}_\nu}
\newcommand{\Gg}{G^\mu{}_\nu}
\newcommand{\Gf}{\widetilde{G}^\mu{}_\nu}
\newcommand{\Tg}{T^\mu{}_\nu}
\newcommand{\Tf}{\widetilde{T}^\mu{}_\nu}
\newcommand{\Teff}{\mathcal{T}^\mu{}_\nu}
\newcommand{\fTeff}{\widetilde{\mathcal{T}}^\mu{}_\nu}

\newcommand{\Sqrt}{S^\mu{}_\nu}

\newcommand{\wt}{\widetilde}
\newcommand{\wb}{\bar}
\newcommand{\Eg}{E^\mu{}_\nu}
\newcommand{\Ef}{\wt{E}^\mu{}_\nu}

\title{\boldmath On the stability of bimetric structure formation}

\author{Marcus Högås, Francesco Torsello and Edvard Mörtsell}

\affiliation{The Oskar Klein Centre, Department of Physics, Stockholm University, SE 106 91, Stockholm, Sweden}

\emailAdd{marcus.hogas@fysik.su.se, francesco.torsello@fysik.su.se, edvard@fysik.su.se}

\abstract{Bimetric gravity can reproduce the accelerated expansion of the Universe, without a cosmological constant. However, the stability of these solutions to linear perturbations has been questioned, suggesting exponential growth of structure in this approximation. We present a simple model of structure formation, for which an analytical solution is derived. The solution is well-behaved, showing that there is no physical instability with respect to these perturbations. The model can yield a growth of structure exhibiting measurable differences from $\Lambda$CDM.}

\begin{document}
\maketitle
\flushbottom

\section{Introduction}
Bimetric gravity (or bimetric relativity) is a theory of a massless and a massive spin-2 field, encoded in two dynamical, interacting metrics \cite{Hassan:2011zd,Hassan:2011ea,Hassan:2012wr,Hassan:2018mbl}. It is a generalization of de Rham--Gabadadze--Tolley (dRGT) massive gravity \cite{deRham:2010ik,deRham:2010kj,Hassan:2011hr} which can be obtained in a certain limit of the bimetric parameter space \cite{Hassan:2014vja}. The dRGT theory is the nonlinear completion of the Fierz--Pauli theory \cite{Fierz:1939ix}. For recent reviews, see \cite{deRham:2014zqa,Schmidt-May:2015vnx}.

Among the virtues of bimetric gravity is that the accelerated expansion of the Universe can be reproduced as a result of the interaction between the metrics; no cosmological constant is needed, see for example \cite{vonStrauss:2011mq,Volkov:2011an,Comelli:2011zm,Akrami:2012vf,Akrami:2013pna,Nersisyan:2015oha}. Also, it may partially address the dark matter phenomenology as a modification of gravity \cite{Enander:2015kda} and provides a dark matter particle candidate in form of a massive graviton \cite{Aoki:2016zgp,Babichev:2016hir,Babichev:2016bxi}.

Unfortunately, all viable background cosmologies with positive ratio of the two scale factors exhibit either a Higuchi ghost or a gradient instability in standard linear cosmological perturbation theory \cite{Comelli:2012db,Khosravi:2012rk,Berg:2012kn,Fasiello:2013woa,Konnig:2014dna,Comelli:2014bqa,DeFelice:2014nja,Solomon:2014dua,Konnig:2014xva,Lagos:2014lca,Enander:2015vja,Konnig:2015lfa,Kenna-Allison:2018izo}. Under some circumstances, the gradient instability emerges when the Hubble parameter exceeds the Fierz--Pauli mass. Taking the general relativity (GR) limit (see e.g. Ref. \cite{Hassan:2014vja}), the instability can be pushed back to an epoch where it has no observable effect \cite{Akrami:2015qga}. The price to pay is that bimetric theory becomes effectively indistinguishable from general relativity in this limit. It has also been suggested that general relativity may be restored nonlinearly via the Vainshtein mechanism when linear perturbation theory breaks down \cite{Vainshtein:1972sx,Aoki:2015xqa,Mortsell:2015exa,Luben:2019yyx}.

There are other ways to deal with this instability however. In linear perturbation theory, the gauge symmetry (general covariance) translates into the freedom of choosing coordinates of the background space-time and coordinates of the perturbed space-time separately (see, e.g., \cite{Weinberg:2008zzc}). In standard linear cosmological perturbation theory, as used to analyze bimetric cosmology, particular choices of variables and gauge are made. The analysis of the Higuchi ghost also relies on a particular gauge choice \cite{Fasiello:2013woa}. Hence, an instability may be an unphysical artifact of a bad choice of variables or gauge. Indeed, as pointed out in Ref. \cite{Ijjas:2018cdm}, standard cosmological perturbation theory is not suited for analyzing the stability of solutions in modified theories of gravity but requires new methods.

Hence, to draw a definite conclusion on the linear stability of bimetric cosmology, one must study linear perturbations in a more general set up, for example along the lines of \cite{Ijjas:2018cdm}. Alternatively, one should solve the full nonlinear equations of motion. For the latter alternative, two routes appear: the first is to derive exact solutions. This will only be possible in special cases. The second is to find numerical solutions. This is a challenging task and partial results have been obtained in spherical symmetry \cite{Kocic:2018ddp,Kocic:2018yvr,cbssn,Kocic:2019zdy,polytropes,meangauge,bimEX}. There is so far no evidence of exponentially growing modes in the numerical evolution of inhomogeneous space-times \cite{polytropes}. Here, we pursue the first option and present an analytical solution of structure formation in bimetric gravity.

\begin{figure}[tbp]
	\centering
	\includegraphics[width=0.75\linewidth]{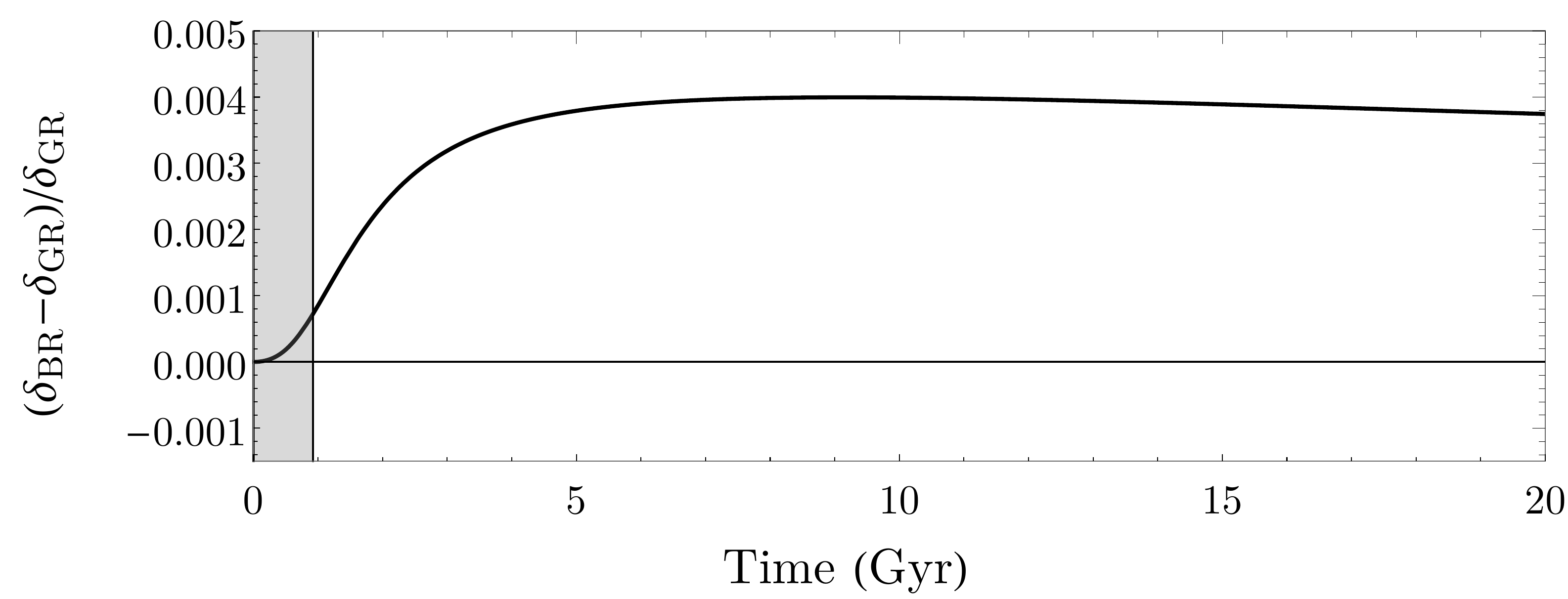}
	\caption{The relative difference of the density contrasts $\delta := (\rho - \wb{\rho})/\wb{\rho}$ in bimetric relativity (BR) and general relativity (GR). The overdensities are homogeneous and set to $\delta = 10^{-5}$ at $z=1100$, approximately corresponding to the time of photon decoupling. Despite $\delta$ being small, note that the results hold nonlinearly. In the shaded region there is an instability in standard cosmological perturbation theory. Here, we have plotted a $\beta_1 \beta_2$ model with parameters chosen according to Refs. \cite{Dhawan:2017leu,Mortsell:2018mfj} to allow for a valid background cosmology. Detailed results and numbers can be found in section \ref{sec:hom_overdens}.}
	\label{fig:GR-vs-BR-structure-formation}
\end{figure}

\paragraph{Summary of results.} The main idea is simple: If one can construct perturbations of bimetric cosmology which solve the complete, nonlinear equations of motion and if these are stable, the background is stable (with respect to these), despite any indications of the contrary at the linear level. In section \ref{sec:hom_overdens}, we derive the full nonlinear solution for a spherically symmetric, homogeneous overdensity. Starting from a Lemaître--Tolman--Bondi Ansatz, the equations of motion implies bimetric Friedmann--Lemaître--Robertson--Walker\footnote{Here, ``bimetric FLRW", refers to homogeneous and isotropic solutions with perfect fluid matter stress--energy.} (FLRW) solutions inside the overdensity, and the solution is unique. Hence, it evolves as a stable homogeneous and isotropic independent universe. Bimetric cosmology is stable with respect to homogeneous overdensities, see for example figure \ref{fig:GR-vs-BR-structure-formation}.

We show that bimetric structure formation can be expected to give corrections to GR ranging from sub-percent level and upwards, see figures \ref{fig:delta_rel-H_rel-beta12} and \ref{fig:delta_rel-H_rel-beta14}. This can be used to exclude certain parameter regimes in the theory, for example with data from the Euclid satellite, being launched in the near future \cite{Laureijs:2011gra,Amendola:2012ys,Amendola:2016saw}.

In section \ref{sec:LTB_hom_lapse}, we study more general perturbations by considering an overdensity of arbitrary radial shape (a bimetric Lemaître--Tolman--Bondi model) and bidiagonal metrics with homogeneous $tt$-components (lapses) in both metrics. We show that this set of assumptions is too constrained to take us beyond a homogeneous overdensity. Hence, an analysis of generic perturbations of bimetric cosmology must invoke numerical solutions of the nonlinear equations of motion.

To enable direct determination of the evolution of a homogeneous overdensity, without the need of numerical calculations, we present a graphical method for evolving the scale factors of the metrics. The method is akin to a particle moving in a potential in Newtonian mechanics.

\paragraph{Notation.} Tildes denote quantities constructed from the metric $f$, otherwise constructed from the metric $g$. Partial derivatives with respect to $t$ are denoted $\partial_t$ or with overdots. Partial derivatives with respect to $r$ are denoted $\partial_r$ or with primes. Overbars denote background cosmology quantities and subscript zero denotes quantities evaluated today. The $B$-parameters are rescaled $\beta$-parameters, $B_n := \kappa_g \beta_n / (\ell^2 H_0^2)$, where $H$ is the Hubble parameter and $\ell$ is a constant length scale determining the Compton wavelength of the massive graviton.

\paragraph{Bimetric gravity.} The equations of motion are,
\begin{subequations}
	\begin{alignat}{2}
		\Eg &:= \Gg - \kappa_g \Teff=0,& \qquad \Teff &:= \Tg + \Vg,\\
		\Ef &:= \Gf - \kappa_f \fTeff=0,& \qquad \fTeff &:= \Tf + \Vf.
	\end{alignat}
\end{subequations}
The Einstein tensors are denoted $\Gg$ and $\Gf$ and the matter stress--energies $\Tg$ and $\Tf$ in the $g$- and $f$-sector, respectively; $\kappa_g$ and $\kappa_f$ are Einstein gravitational constants of the two metrics. From now on, we set $\Tf=0$. The bimetric stress--energies, $\Vg$ and $\Vf$, are constructed to avoid the Boulware--Deser ghost plaguing generic theories of massive gravity \cite{Boulware:1973my},
\begin{subequations}
	\begin{align}
	\Vg &:= - \frac{1}{\ell^2} \sum_{n=0}^{3} \beta_n \sum_{k=0}^{n} (-1)^{n+k} e_k(S) {(S^{n-k})}^\mu{}_\nu,\\
	\Vf &:= - \frac{1}{\ell^2} \sum_{n=0}^{3}\beta_{4-n} \sum_{k=0}^{n} (-1)^{n+k} e_k(S^{-1}) {(S^{-n+k})}^\mu{}_\nu,
	\end{align}
\end{subequations}
with $e_n(S)$ being the elementary symmetric polynomials of the principal square root $S := (g^{-1}f)^{1/2}$ \cite{Hassan:2017ugh}, $\beta_n$ dimensionless constants. If nothing else is stated, we use geometrized units in which Newton's gravitational constant and the speed of light are set to one ($G_\mathrm{N}=c=1$). In these units, there appears in the equations a length scale $\ell$ which parameterizes the Compton wavelength of the massive graviton, whereas in so-called natural units ($c = \hbar =1$), a mass scale $m$ parameterizing the graviton mass appears. The bimetric conservation law and conservation of stress--energy read, respectively,
\begin{equation}
\label{eq:conservations}
	\nabla_\mu \Vg =0, \quad \nabla_\mu \Tg =0.
\end{equation}
The former equation follows from the assumption of matter stress--energy conservation. The conservation equation $\wt{\nabla}_\mu \Vf = 0$ follows from the contracted Bianchi identity of $f$.

\section{Structure formation}
\subsection{Evolution of a homogeneous overdensity}
\label{sec:hom_overdens}
In general relativity, the Lemaître--Tolman--Bondi (LTB) model provides an analytical solution of an arbitrary, spherically symmetric distribution of massive, pressureless, dust particles \cite{Lemaitre:1933,Tolman:1934,Bondi:1947fta}. For example, we can specify the initial matter distribution to be almost homogeneous in which case we have a spherically symmetric perturbation of a homogeneous and isotropic universe filled with dust (or during matter domination).\footnote{As another example, a homogeneous, spherical dust cloud with vacuum outside can be used as a simple model of the gravitational collapse of a star. This is the Oppenheimer--Snyder space-time \cite{Oppenheimer:1939ue}. In GR, this implies FLRW inside the cloud and Schwarzschild outside, due to Birkhoff's theorem \cite{Jebsen:1921,Jebsen2005,Birkhoff:1923}. In bimetric theory, there is no such statement \cite{Kocic:2017hve,polytropes}.}

The LTB model is an exact solution to the nonlinear Einstein equations. In bimetric theory, an analog model would provide useful input into the stability of bimetric cosmology. The only such candidate in the literature would be the Vaidya-like solution of collapsing, massless particles \cite{Hogas:2019cpg}. However, it is not an overdensity of a standard cosmological background. In the following section, we derive the evolution of a homogeneous overdensity and show that a bimetric LTB model necessarily reduces to a bimetric FLRW solution, given the assumptions below. Thereafter, we use it as a simple model of structure formation.

We start with a general bidiagonal Ansatz for the metrics with the same spherical symmetry in both sectors \cite{Torsello:2017ouh},
\begin{subequations}
	\label{eq:Ansatz}
	\begin{align}
		ds_g^2 &= - d t^2 + A^2 (t,r) d r^2 + B^2(t,r) r^2 d \Omega^2, \qquad \qquad d \Omega^2 := d \theta^2 + \sin^2 \theta d \phi^2,\\
		ds_f^2 &= - \wt{\alpha}^2(t,r) d t^2 + \wt{A}^2 (t,r) d r^2 + \wt{B}^2(t,r) r^2 d \Omega^2.
	\end{align}
\end{subequations}
It follows trivially that the principal square root is (i.e. with positive real part of the eigenvalues \cite{higham2008functions}),
\begin{equation}
	\Sqrt = \mathrm{diag}(\wt{\alpha},\wt{A}/A,\wt{B}/B,\wt{B}/B),
\end{equation}
where we have assumed $\wt{\alpha},A,B,\wt{A},\wt{B}>0$ without loss of generality. We have chosen the coordinates to be comoving with the dust particles, that is,
\begin{equation}
\label{eq:Tgdust}
	\Tg = \mathrm{diag}(-\rho(t,r),0,0,0).
\end{equation}
From the $E^t{}_r$ equation it follows that,
\begin{equation}
\label{eq:Asol}
	A(t,r) = \frac{[rB(t,r)]'}{\sqrt{1-k(r)}},
\end{equation}
where $k(r)$ is a generic function of $r$ (and $<1$). Conservation of stress--energy now reads,
\begin{equation}
\label{eq:stressenCons}
	\partial_t \left[\rho B^2 \left(r B\right)'\right] =0. 
\end{equation}
Thus, $\rho B^2 (rB)'$ is a conserved quantity. With a general $\rho(t,r)$, it does not seem to be any straightforward way to proceed and obtain an analytical solution. However, let us focus on the case of a homogeneous overdensity where $\rho$ is constant as a function of $r$ within some radius $r_*$,
\begin{equation}
\rho(t,r) = \rho(t), \quad r < r_*.
\end{equation}
Equation \eqref{eq:stressenCons} implies,
\begin{equation}
\label{eq:conserv}
	\rho(t) B^2(t,r) \left(rB(t,r)\right)' = \left. \rho(t) B^2(t,r) \left(rB(t,r)\right)' \right|_{t=0} = \rho(0), \quad r<r_*,
\end{equation}
where the gauge freedom was used to set $B(0,r)=1$. Integrating \eqref{eq:conserv},
\begin{equation}
	B^3(t,r) = \frac{\rho(0)}{\rho(t)} + \frac{b(t)}{r^3}, \quad r<r_*,
\end{equation}
where $b(t)$ is an arbitrary function of $t$. Regularity at $r=0$ requires $B$ to be either even or odd as a function of $r$, hence $b=0$ \cite{Ruiz:2007rs}. Thus, $B(t,r)$ is a function of $t$ only and we can define the scale factor $a$ as,
\begin{equation}
	a^3(t) := \rho(0)/\rho(t) \quad \Rightarrow \quad B(t,r) = a(t), \quad A(t,r) = \frac{a(t)}{\sqrt{1-Kr^2}}, \quad r<r_*,
\end{equation}
with $K$ constant. Hence, the region $r<r_*$ is described by a bimetric FLRW model. In other words, a homogeneous overdensity behaves as a mini-universe, like in GR. Note that this result is non-trivial since the (local) homogeneity of $\rho$ does not automatically imply homogeneity of $g$ \cite{Torsello:2017ouh}. 

\begin{figure}[t]
	\centering
	\includegraphics[width=1.0\linewidth]{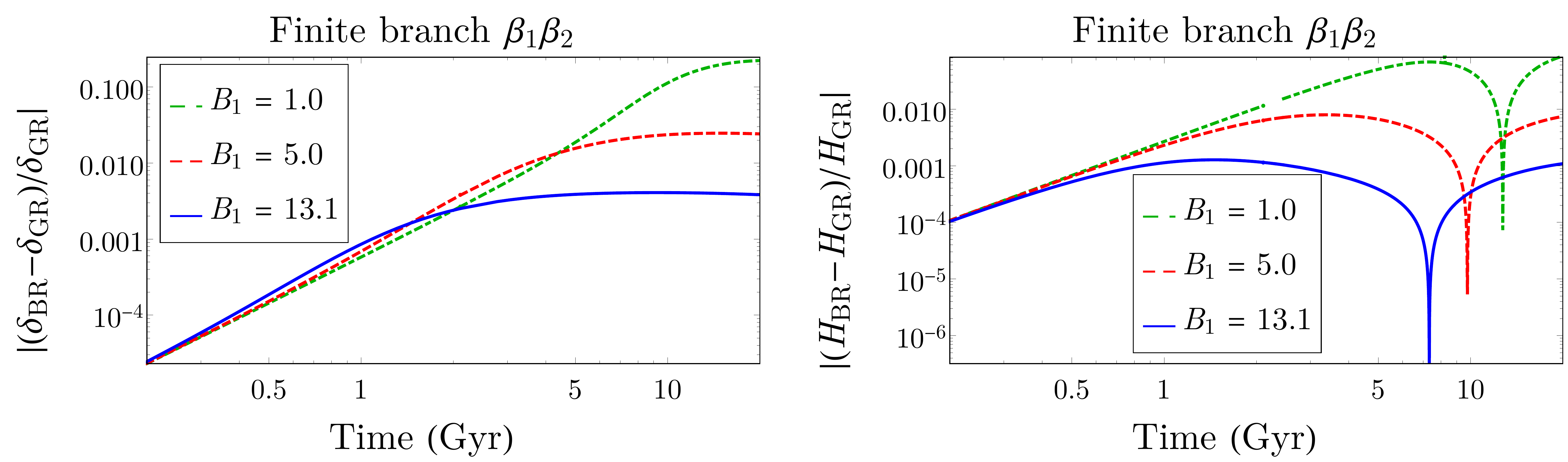}
	\caption{Finite branch $\beta_1 \beta_2$ models. \textit{Left panel}: The relative difference of the density contrasts $\delta$ in bimetric relativity (BR) and general relativity (GR). This result should be compared with those of standard cosmological perturbation theory exhibiting a gradient instability, see for example figure 3 of \cite{Solomon:2014dua}. The overdensities have a stable evolution and the results hold nonlinearly. \textit{Right panel}: The relative difference of the background Hubble parameters in BR and GR.}
	\label{fig:delta_rel-H_rel-beta12}
\end{figure}

\begin{figure}[h]
	\centering
	\includegraphics[width=1.0\linewidth]{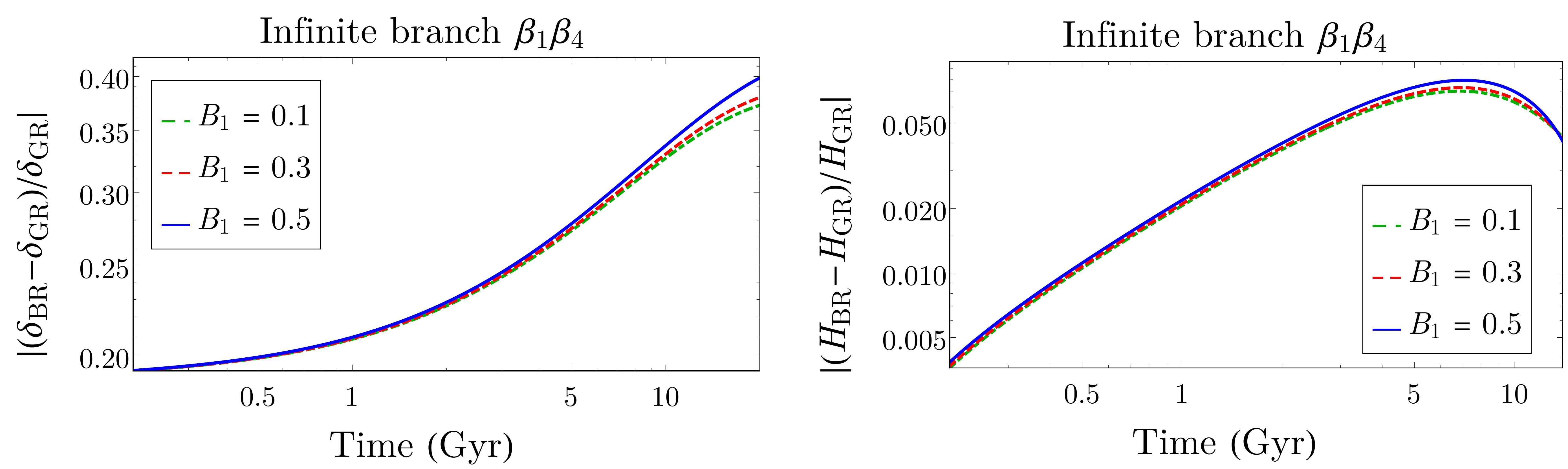}
	\caption{Results for infinite branch $\beta_1 \beta_4$ models. The overdensities have a stable evolution and the results hold nonlinearly.}
	\label{fig:delta_rel-H_rel-beta14}
\end{figure}

It is now straightforward to show that the background is stable with respect to such perturbations: We start by defining the density contrast $\delta := (\rho - \wb{\rho})/\wb{\rho}$ which quantifies the magnitude of the perturbation relative to the background. As can be seen in figures \ref{fig:delta_rel-H_rel-beta12} and \ref{fig:delta_rel-H_rel-beta14}, the difference in $\delta$ between the bimetric and the corresponding GR perturbation of a $\Lambda$CDM model stays bounded. The bimetric model must be stable since the GR model is. As examples of structure formation in bimetric gravity, we look at perturbations of two types of background cosmologies which can be made compatible with observations---the finite branch $\beta_1 \beta_2$ models and the infinite branch $\beta_1 \beta_4$ models, see appendix \ref{sec:BRcosmo}. As in GR, due to the simplicity of a homogeneous overdensity, the results should not be regarded as accurate predictions of structure formation. Rather, we use them to estimate the difference in the growth of structure between GR and bimetric gravity.

In both examples, we start with an initial overdensity of size $\delta = 10^{-5}$ at redshift $z = a_0/a-1 = 1100$, approximately corresponding to the time of photon decoupling \cite{Smoot:1992td}. It should be stressed that even if we take the initial overdensity to be small, the model is still a solution to the full, nonlinear equations of motion. The matter stress--energy is assumed to consist of pressureless dust. One of the $B_n$ can be expressed in terms of the other and the background matter density $\Omega_{\mathrm{M},0}$, as explained in appendix \ref{sec:BRcosmo}. The Hubble parameter today is set to $H_0 = 70\, \mathrm{km/s/Mpc}$ and $\kappa_g / \kappa_f = 1$ to enable comparison with most bimetric cosmologies. The growth rate $f_g$ and growth index $\gamma$ are defined as,
\begin{equation}
\label{eq:fg_fit}
	f_g := \frac{d \log \delta}{d \log a} \simeq \left[\left(\frac{H_0}{H}\right)^2 \Omega_\mathrm{M}\right]^\gamma,
\end{equation}
where $\gamma$ is the best fit value of the right-hand side expression of \eqref{eq:fg_fit} to the left-hand side of \eqref{eq:fg_fit}.

\begin{figure}[h]
	\centering
	\includegraphics[width=0.5\linewidth]{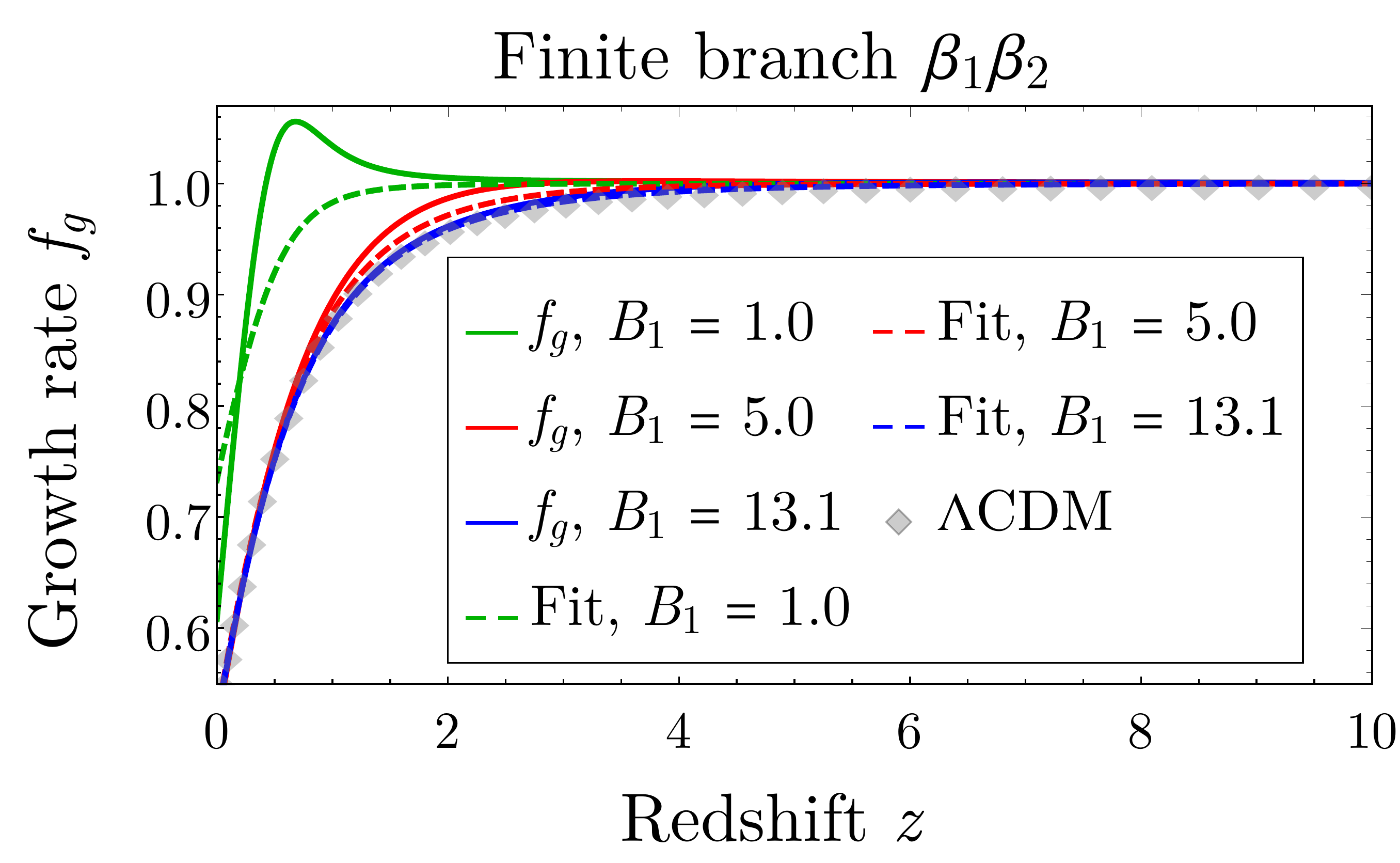}
	\caption{Growth rate $f_g$ for a homogeneous overdensity and fitted function, given by $(H_0^2 \Omega_{\mathrm{M}} / H^2)^\gamma$. Finite branch $\beta_1 \beta_2$ models and GR reference ($\Lambda$CDM).}
	\label{fig:growth-rate-beta12}
\end{figure}

\paragraph{GR reference.} As a reference solution, we take a GR $\Lambda$CDM model with $\Omega_{\mathrm{M},0}=0.30$, $\Omega_{\Lambda,0} = 0.70$. \newpage

\paragraph{Finite branch $\beta_1 \beta_2$ models.} Here, the matter density today is set to $\Omega_{\mathrm{M},0}=0.3$. We consider three cases: $B_1 = 1.0$, $B_1 = 5.0$, and $B_1 = 13.1$. The latter model corresponds to a valid cosmology at the level of the background \cite{Dhawan:2017leu,Mortsell:2018mfj}. For this model, $\delta$ deviates at most $\simeq 0.4 \, \%$ from GR whereas for smaller $B_1$ the deviation reaches above $\simeq 20 \, \%$, as can be seen in figure \ref{fig:delta_rel-H_rel-beta12}. Hence, bimetric structure formation can be expected to give sub-percent corrections to GR for a $\beta_1 \beta_2$ model consistent with background data. However, the $B_1 = 1.0$ and $B_1 = 5.0$ models are already excluded at level of background cosmology \cite{Dhawan:2017leu,Mortsell:2018mfj}. In figure \ref{fig:delta_rel-H_rel-beta12}, we also plot the relative difference of the Hubble parameters of the background cosmologies in bimetric and general relativity. The spikes are due to $H_\mathrm{BR}-H_\mathrm{GR}$ crossing zero, as could be understood from figure \ref{fig:graphical-sol}.\footnote{In figure \ref{fig:graphical-sol}, we have normalized the scale factors so that $a=1$ corresponds to $H_\mathrm{BR} = H_\mathrm{GR} = 70 \,\mathrm{km/s/Mpc}$. Note that the age of the Universe is different in bimetric gravity and GR, so $a$ is not equal to one at the same point in time in the two theories. Hence the shift in the peaks of figure \ref{fig:delta_rel-H_rel-beta12}.} The crossing is smooth; the apparent sharpness in the figure is due to the log scale. Comparing the left and right panels of figure \ref{fig:delta_rel-H_rel-beta12}, the difference in the Hubble parameters for the background cosmologies (right panel) is of the same order of magnitude as the difference in density contrasts (left panel), at least if the Hubble parameter crossing region is  excluded.

\noindent In figure \ref{fig:growth-rate-beta12}, the bimetric growth rate is plotted in the range $0<z<10$ together with the fitted function $(H_0^2 \Omega_{\mathrm{M}} / H^2)^\gamma$ and the $\Lambda$CDM reference. The $B_1 = 13.1$ model follows GR closely and has a good fit, whereas the $B_1=1.0$ model deviates significantly and the fit is poor.

\begin{figure}[h]
	\centering
	\includegraphics[width=0.5\linewidth]{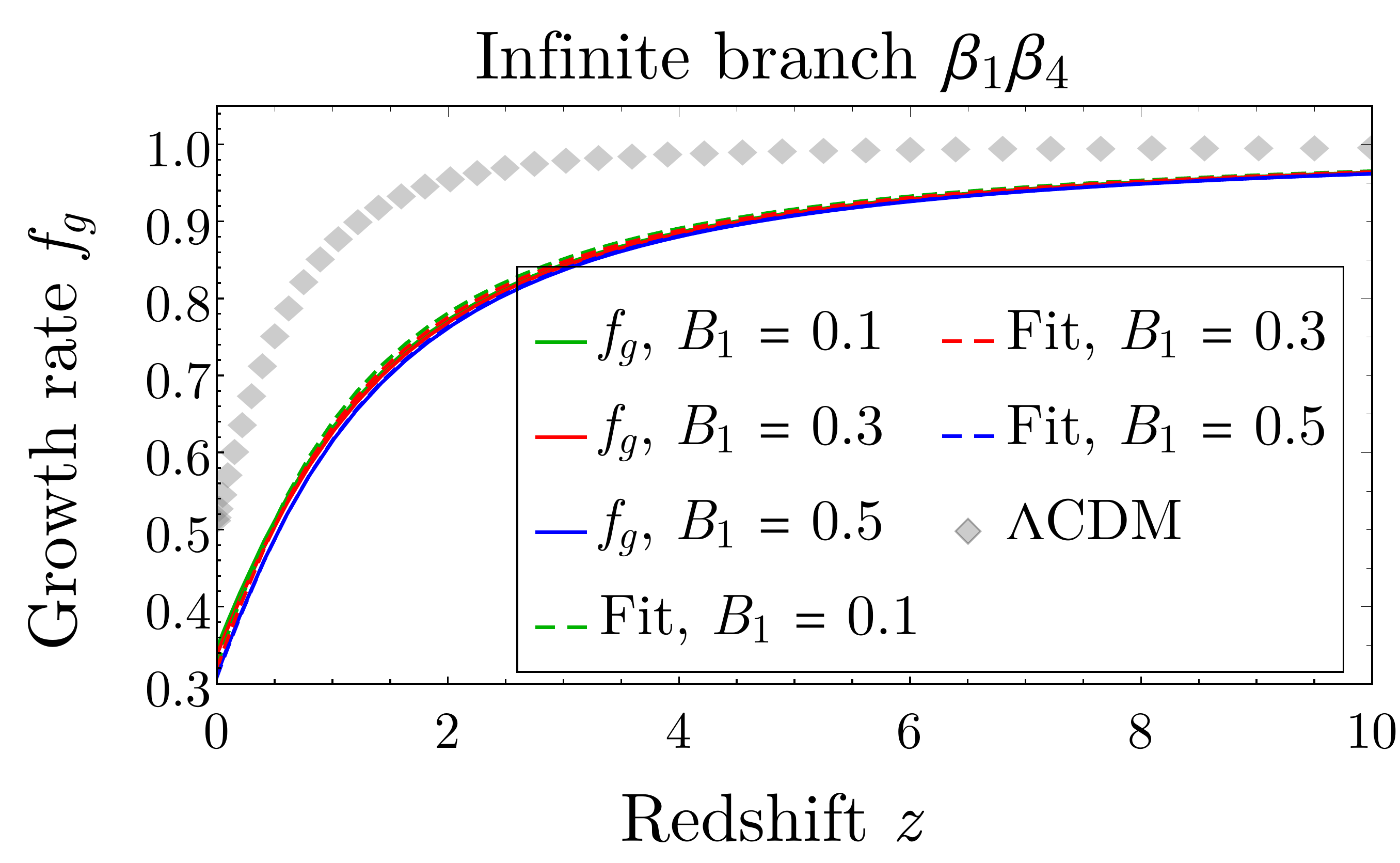}
	\caption{Growth rate $f_g$ for a homogeneous overdensity and fitted function $(H_0^2 \Omega_{\mathrm{M}} / H^2)^\gamma$. Infinite branch $\beta_1 \beta_4$ models and GR reference ($\Lambda$CDM).}
	\label{fig:growth-rate-beta14}
\end{figure}

\paragraph{Infinite branch $\beta_1 \beta_4$ models.} Here, theoretical and experimental observations require a $B_1$ in the range $0<B_1<0.529$. The best fit model to supernovae observations is given by $\Omega_{\mathrm{M},0} \simeq 0.16$, which we adopt here \cite{Konnig:2013gxa}. The density contrast $\delta$ grows apart from GR to reach a maximum around $\simeq 40 \, \%$ at late times, see figure \ref{fig:delta_rel-H_rel-beta14}. As opposed to the $\beta_1 \beta_2$ models, the difference in structure formation from GR (figure \ref{fig:delta_rel-H_rel-beta14}, left panel) is an order of magnitude greater than the difference in background cosmologies (figure \ref{fig:delta_rel-H_rel-beta14}, right panel). Thus, structure formation data could be used to exclude these $\beta_1 \beta_4$ models with consistent background expansions. In figure \ref{fig:growth-rate-beta14}, the bimetric growth rate and fitted $(H_0^2 \Omega_{\mathrm{M}} / H^2)^\gamma$ are plotted in the range $0<z<10$ for different models, including GR ($\Lambda$CDM) as a reference. The fits follow $f_g$ closely, but they all deviate significantly from the GR reference.

The calculated growth index in the range $0 < z < 1100$ is shown in table \ref{table:gamma}. Since they are stable, our results can be trusted, unlike the linear results which suffer from either Higuchi or gradient instabilities. The curious $\gamma$ value of the $B_1 = 1.0$ model can be understood as the fitted function departs heavily from the growth rate $f_g$, as seen in figure \ref{fig:growth-rate-beta12}. The Euclid satellite is expected to distinguish a $\gamma$ which deviates more than $\simeq 0.03$ from the GR value \cite{Laureijs:2011gra,Amendola:2012ys,Amendola:2016saw}. Thus, it should be possible to distinguish between GR and the infinite branch $\beta_1 \beta_4$ models as well as some regions in the parameter space of finite branch $\beta_1 \beta_2$ models.

It should be noted that the resulting value of $\gamma$ depends somewhat on the choice of initial conditions, see appendix \ref{sec:BRstruct}. Nevertheless, the relative differences between the GR and bimetric values remain essentially the same.

\begin{table}[t]
	\caption{Growth index for a homogeneous overdensity.}
	\centering
	\begin{tabular}{r | c | c | c | c | c | c | c}
		\hline\hline
		Model: & $\beta_1 \beta_2$ & $\beta_1 \beta_2$ & $\beta_1 \beta_2$ & $\beta_1 \beta_4$ & $\beta_1 \beta_4$ & $\beta_1 \beta_4$ & $\Lambda$CDM (GR)\\
		\hline $B_1 = $ & 1.0 & 5.0 & 13.1 & 0.1 & 0.3 & 0.5 &---\\
		$\gamma = $ & 0.258 & 0.544 & 0.546 & 0.615 & 0.623 & 0.641 & 0.542\\[0.5ex]
		\hline 
	\end{tabular}
	\label{table:gamma}
\end{table}

\noindent To enable a direct determination of the evolution of an overdensity, without the need of numerical integration, we construct a graphical method of solving for the scale factors $a(t):=B$ and $\widetilde{a}(t):=\widetilde{B}$, see appendix \ref{sec:graphical}.

Concluding the section:\\

\noindent \emph{Assuming bidiagonal and spherically symmetric metrics, the bimetric FLRW model is the unique solution inside a homogeneous, pressureless dust cloud. Hence, there is no instability of bimetric cosmology with respect to spherically symmetric, homogeneous overdensities. The solution can be used as a simple model of structure formation, predicting, for some parameter choices, deviations from $\mathit{\Lambda}$CDM that will be detectable, using future data from the Euclid satellite.}\\

\subsection{Generalization: Lemaître--Tolman--Bondi model}
\label{sec:LTB_hom_lapse}
In section \ref{sec:hom_overdens}, we analyzed a restricted model, assuming a homogeneous matter overdensity. A bimetric analog of the LTB model in GR would describe the evolution of arbitrary spherical distributions of $\rho$. One such model was proposed in Ref. \cite{Nersisyan:2015oha}. Here, we perform a detailed analysis, showing that the assumption of an arbitrary radial dependence of $\rho$ cannot be maintained in this set up. In fact, homogeneity is required and bimetric FLRW recovered.

The Ansatz is, 
\begin{subequations}
	\label{eq:hom_lapse_Ansatz}
	\begin{align}
	ds_g^2 &= - d t^2 + A^2(t,r) d r^2 +B^2 (t,r) r^2 d \Omega^2,\\
	ds_f^2 &= - \wt{\alpha}^2(t) d t^2 + \wt{A}^2(t,r) d r^2 + \wt{B}^2(t,r) r^2 d \Omega
	^2.
	\end{align}
\end{subequations}
The assumption of a homogeneous lapse in the $f$-sector, $\wt{\alpha}(t)$, calls for attention since it must be ensured that this assumption is not violated. 

As in the model of a homogeneous overdensity (section \ref{sec:hom_overdens}), the $E^t{}_r$ equation implies \eqref{eq:Asol}. Similarly, with a homogeneous lapse $\widetilde{\alpha}(t)$, the $\wt{E}^t{}_r$ equations can be integrated to,
\begin{equation}
\label{eq:AtoB}
\wt{A}(t,r) = \frac{[r \wt{B}(t,r)]'}{\sqrt{1-\wt{k}(r)}},
\end{equation}
with $\wt{k}(r)$ arbitrary ($<1$). The $r$-component of the bimetric conservation law \eqref{eq:conservations} reads,
\begin{equation}
\label{eq:rBRcons}
\left(\sqrt{1-k \vphantom{\widetilde{k}}}-\sqrt{1-\wt{k}}\right) \left[B(\beta_1 + \beta_2 \wt{\alpha}) + \wt{B}(\beta_2 + \beta_3 \wt{\alpha})\right]= 0.
\end{equation}
Setting the second parenthesis to zero implies, after some calculations,
\begin{equation}
\label{eq:gkIsfk}
k(r) = \wt{k}(r),
\end{equation}
for a non-static solution. This is also the solution when setting the first parenthesis to zero. Thus, $k = \wt{k}$ is the unique solution of \eqref{eq:rBRcons}. The $t$-component of the bimetric conservation law \eqref{eq:conservations} gives,
\begin{equation}
\wt{\alpha} = \frac{\partial_r \left(r^3 \, \mathcal{F} \, \dot{\wt{B}}\right)}{\partial_r \left(r^3 \, \mathcal{F} \, \dot{B}\right)}, \quad \mathcal{F} := \beta_1 B^2 + 2 \beta_2 B \wt{B}+ \beta_3 \wt{B}^2.
\end{equation}
Integrating with respect to $r$,
\begin{equation}
\label{eq:falpha}
\wt{\alpha}(t) = \frac{\dot{\wt{B}}(t,r)}{\dot{B}(t,r)},
\end{equation}
where a free function of $t$ is set to zero to ensure regularity. From \eqref{eq:AtoB}, \eqref{eq:gkIsfk}, and \eqref{eq:falpha}, it follows that the bimetric equations of motion take the following form (for details, see appendix \ref{sec:details}),
\begin{subequations}
	\label{eq:LTBeqns}
	\begin{align}
	\left(\frac{\dot{B}}{B}\right)^2 + \frac{k}{r^2 B^2} -\kappa_g \frac{M}{r^3 B^3} - \frac{\kappa_g}{\ell^2} \left(\frac{\beta_0}{3} + \beta_1 y + \beta_2 y^2 + \frac{\beta_3}{3}y^3\right)&= 0,\\
	\label{eq:y_poly_all_eqns}
	\frac{1}{\ell^2} \left[-\frac{\beta_3}{3} y^3 + \left(\frac{\beta_4}{3\kappa} - \beta_2\right)y^2+ \left(\frac{\beta_3}{\kappa}- \beta_1 \right)y + \left(\frac{\beta_2}{\kappa}- \frac{\beta_0}{3}\right) + \frac{\beta_1}{3\kappa} y^{-1}\right] -\frac{M}{r^3 B^3} &= 0,
	\end{align}
\end{subequations}
in addition to equations \eqref{eq:stressenCons} and \eqref{eq:falpha}. Here, $\kappa := \kappa_g / \kappa_f$. Note the similarity between \eqref{eq:LTBeqns} and the bimetric FLRW equations \eqref{eq:BRcosmo}. The difference is that here, the fields still depend on both $t$ and $r$.

Recalling $y := \wt{B}/B$, it follows that,
\begin{equation}
\label{eq:flapse}
\wt{\alpha} = \frac{\partial \widetilde{B} / \partial t }{\partial B / \partial t} = \frac{\partial \widetilde{B}}{\partial B} = \frac{\partial (y B)}{\partial B} = y + B \frac{\partial y}{\partial B}.
\end{equation}
Solving \eqref{eq:y_poly_all_eqns} for $y$, plugging the solution into the right-hand side of \eqref{eq:flapse}, differentiating both sides with respect to $r$ and setting the result to zero (remember $\partial_r \wt{\alpha}=0$), we obtain a differential equation for $B$ which can be solved with the result,
\begin{equation}
\label{eq:B_separable}
B(t,r) = a(t) b(r),
\end{equation}
for some functions $a$ and $b$. From this, the bimetric FLRW Ansatz is finally recovered (after a few steps, given in appendix \ref{sec:details}),
\begin{subequations}
	\label{eq:bimetricFLRW}
	\begin{alignat}{2}
	ds_g^2 &= - d t^2 + a(t)^2 \left(\frac{d r^2}{1-K r^2} + r^2 d \Omega^2\right),& \quad K &= \mathrm{const.},\\
	ds_f^2 &= - \wt{\alpha}^2(t)d t^2 + \wt{a}(t)^2 \left(\frac{d r^2}{1-K r^2} + r^2 d \Omega^2\right),& \quad \wt{\alpha}(t) &= \frac{\dot{\wt{a}}(t)}{\dot{a}(t)} .
	\end{alignat}
\end{subequations}
\newpage 

\noindent We summarize:\\

\noindent \emph{Starting from a bidiagonal Ansatz with the same spherical symmetry in both sectors, a homogeneous lapse in $f$ \eqref{eq:hom_lapse_Ansatz}, and pressureless dust stress--energy, the model necessarily reduces to bimetric FLRW with a homogeneous $\rho$.}\\

\noindent To obtain a bimetric LTB model where the density profile can be freely specified at some initial point in time, one must go beyond the Ansatz \eqref{eq:hom_lapse_Ansatz}. For example, allowing for a radial dependence in the $f$ lapse $\wt{\alpha}$ or adding off-diagonal ($d t d r$) terms to the $f$ metric. In that case, only numerical methods are available.

As a final note, to obtain a global solution, the inside space-time should be matched to an outside solution. Assuming, naïvely, that the junction conditions in GR are simply doubled in bimetric theory, one requires the induced metrics and extrinsic curvatures to be continuous over the boundary surface. Implementing these constraints, one can show that the area radii of the two metrics ($rB$ and $r \widetilde{B}$) must not be proportional on the boundary surface. This rules out all the static, spherically symmetric, vacuum space-times as outside solutions. Hence, an inside bimetric FLRW solution cannot be matched with a static, spherically symmetric, vacuum space-time outside. This can be seen as a manifestation of the violation of Birkhoff's theorem in bimetric gravity; the outside vacuum solution must be nonstatic. The nonstatic vacuum solution of \cite{Kocic:2017hve} is ruled out as an outside space-time due to a pathological bimetric FLRW solution for the choice of parameters therein, corresponding to the partially massless parameters with only $\beta_{0,2,4}$ nonzero.

\section{Summary and outlook}
In section \ref{sec:hom_overdens}, we derived the analytical solution for a spherically symmetric, homogeneous overdensity. The solution is unique and describes a bimetric Friedmann--Lemaître--Robertson--Walker mini-universe inside the overdensity. As examples we considered overdensities on finite branch $\beta_1 \beta_2$ and infinite branch $\beta_1 \beta_4$ cosmologies, including a valid $\beta_1 \beta_2$ model for the background cosmology. The deviations in structure formation from general relativity ranges from sub-percent level for the valid background $\beta_1 \beta_2$ model to $\simeq 40\, \%$, see figures \ref{fig:delta_rel-H_rel-beta12} and \ref{fig:delta_rel-H_rel-beta14}. 

In section \ref{sec:LTB_hom_lapse}, we showed that it is necessary to allow for a radial dependence in the $f$ lapse ($tt$-component) or to add off-diagonal terms to the $f$ metric if one wants to solve the full nonlinear equations of motion for a spherically symmetric overdensity with an arbitrary radial profile. Unfortunately, there is no simple bimetric analog of the Lemaître--Tolman--Bondi solutions in general relativity. This displays the importance of numerical bimetric relativity. 

The analytical solution of section \ref{sec:hom_overdens} shows that bimetric cosmology is stable with respect to such perturbations. The result suggests that perturbations around the bimetric cosmologies should be revisited in a more general context where, for example, the gauge (coordinates) is not restricted to a particular choice. It may be possible to show that the instabilities are mere gauge effects. Another prospect is to analyze whether there exists a cosmological Vainshtein mechanism that restores general relativity in the regions of linear instability \cite{Aoki:2015xqa,Mortsell:2015exa,Luben:2019yyx}. The results of this paper is compatible with such a scenario. Until these options have been explored, there is no conclusive reason to think that bimetric cosmology is unstable with respect to perturbations of homogeneous and isotropic backgrounds.

\acknowledgments
Thanks to Mikica Kocic for many fruitful discussions on the subject and to Marvin Lüben, Angelo Caravano, and an anonymous referee for useful comments.

\appendix
\section{Bimetric cosmology}
\label{sec:BRcosmo}
Assuming the same homogeneity and isotropy in both sectors, the metrics read,
\begin{subequations}
	\label{eq:BRFLRWAnsatz}
	\begin{align}
		ds_g^2 &= - d t^2 + a^2(t) \left(\frac{d r^2}{1-K r^2} + r^2 d \Omega^2\right), \quad K = \mathrm{const.},\\
		ds_f^2 &= - \wt{\alpha}^2(t)d t^2 + \wt{a}^2(t) \left(\frac{d r^2}{1-K r^2} + r^2 d \Omega^2\right),
	\end{align}
\end{subequations}
in comoving coordinates. For pressureless dust, conservation  of stress--energy $\Tg = \mathrm{diag}(-\rho(t),0,0,0)$ leads to,
\begin{equation}
\label{eq:rho2a}
	\rho(t) = \rho_0 \left[a_0 / a(t)\right]^3.
\end{equation}
With the Ansatz \eqref{eq:BRFLRWAnsatz}, the bimetric conservation law \eqref{eq:conservations} is,
\begin{equation}
	\left[\beta_1 +2 \beta_2 y + \beta_3 y^2 \right]\left(\dot{\wt{a}}- \dot{a} \wt{\alpha}\right)=0, \quad y := \wt{a}/a,
\end{equation}
and thus have two branches of solutions. Setting the first parenthesis to zero yields the algebraic branch where $y$ is expressed in terms of the constant $\beta$-parameters. This branch is usually discarded due to possibly problematic perturbations \cite{Schmidt-May:2015vnx}. Setting the second parenthesis to zero defines the dynamical branch in which,
\begin{equation}
\label{eq:FLRWlapse}
	\wt{\alpha} = \dot{\wt{a}} / \dot{a}.
\end{equation}
Using \eqref{eq:rho2a} and \eqref{eq:FLRWlapse}, the equations of motion are,
\begin{subequations}
	\label{eq:BRcosmo}
	\begin{align}
	\label{eq:BRFriedmann}
		\frac{H^2}{H_0^2} - \Omega_{\mathrm{M}}(a) - \Omega_{\mathrm{DE}}(a) - \Omega_K(a) &=0,\\
		\label{eq:ypoly}
		- \frac{B_1}{3\kappa} y^{-1} + \left(\frac{B_0}{3} - \frac{B_2}{\kappa}\right) + \left(B_1 - \frac{B_3}{\kappa} \right) y + \left(B_2 - \frac{B_4}{3\kappa}\right)y^2 + \frac{B_3}{3} y^3 +  \Omega_{\mathrm{M}}(a) &= 0,
	\end{align}
\end{subequations}
with,
\begin{alignat}{3}
\label{eq:cosmodefs}
	H &:= \frac{\dot{a}}{a},& \quad \Omega_{\mathrm{M}}(a) &:= \frac{\kappa_g}{3 H_0^2} \rho,& \quad \Omega_{\mathrm{DE}}(a) &:= \frac{B_0}{3} + B_1 y +B_2 y^2 +\frac{B_3}{3} y^3,\nonumber\\
	\Omega_K(a) &:= \frac{-K}{H_0^2 a^2},&  \quad B_n &:= \frac{\kappa_g \beta_n}{\ell^2 H_0^2}.
\end{alignat}
The energy densities depend on the scale factor and $\Omega_{\mathrm{M}}$ is measured in terms of the critical density today. Equation \eqref{eq:ypoly} is a quartic polynomial in $y$ and can thus in principle be solved analytically in terms of $\Omega_{\mathrm{M}}$ and the $\beta$-parameters. Doing so, there may be several branches of real solutions. The cosmological models can be classified according to the behavior of $y$ at early times. If $y \gg 1$ when $\Omega_{\mathrm{M}} \gg 1$, they are referred to as infinite branch solutions. If $y \ll 1$ when $\Omega_{\mathrm{M}} \gg 1$, they are referred to as finite branch solutions.

\noindent Applying standard linear cosmological perturbation theory on a $\beta_1 \beta_2$ model, assuming pressureless dust in the matter stress--energy, there appears a gradient instability for sub-horizon scalar modes at times earlier than the point in time at which,
\begin{align}
	18 B_2 \left(\kappa B_1^2 + 4 B_2^2 \right) y^5 + 9 B_1 \left(\kappa B_1^2 + 10 B_2^2 \right) y^4 +48 B_1^2 B_2 y^3 +& \nonumber\\
	+6 B_2 \left(2 B_1^2 -\kappa^{-1} B_2^2 \right) y^2 -6 \kappa^{-1} B_1^2 B_2 y-\kappa^{-1} B_1^3&=0,
\end{align}
see for example \cite{Akrami:2015qga}.

In this paper we analyze structure formation for two-parameter models. That is, with precisely two nonzero $\beta$-parameters. For example, with only $\beta_1$ and $\beta_2$ nonzero. A two-parameter model effectively gives room to adjust one of them freely. The other one is determined as follows. Evaluating the modified Friedmann equation \eqref{eq:BRFriedmann} today, with $K=0$,
\begin{equation}
\label{eq:OmegasId}
	\Omega_{\mathrm{M},0} + \Omega_{\mathrm{DE},0} =1.
\end{equation}
From the definition of $\Omega_{\mathrm{DE}}$ \eqref{eq:cosmodefs}, one can solve for $y$ in terms of $\Omega_{\mathrm{DE}}$. This solution can be plugged into the quartic polynomial for $y$ \eqref{eq:ypoly}, yielding an equation in $\Omega_{\mathrm{M}}$, $\Omega_{\mathrm{DE}}$ and the $\beta$-parameters. Evaluating today and using \eqref{eq:OmegasId} gives an equation for one $\beta$ in terms of the other and $\Omega_{\mathrm{M},0}$. Imposing a value of $\Omega_{\mathrm{M},0}$ gives a relation between the two $\beta$. For the $\beta_1 \beta_2$ model,
\begin{equation}
	B_2 = \frac{-B_1^2 +9 (1-\Omega_{\mathrm{M},0}) - \sqrt{B_1^4 + 9 B_1^2 (1-\Omega_{\mathrm{M},0})}}{9 (1-\Omega_{\mathrm{M},0})},
\end{equation} 
and for the $\beta_1 \beta_4$ model,
\begin{equation}
	B_4 = \frac{3 (1-\Omega_{\mathrm{M},0}) B_1^2 - B_1^4}{(1-\Omega_{\mathrm{M},0})^3}.
\end{equation}

\section{Structure formation}
\label{sec:BRstruct}
\begin{figure}[tbp]
	\centering
	\includegraphics[width=0.47\linewidth]{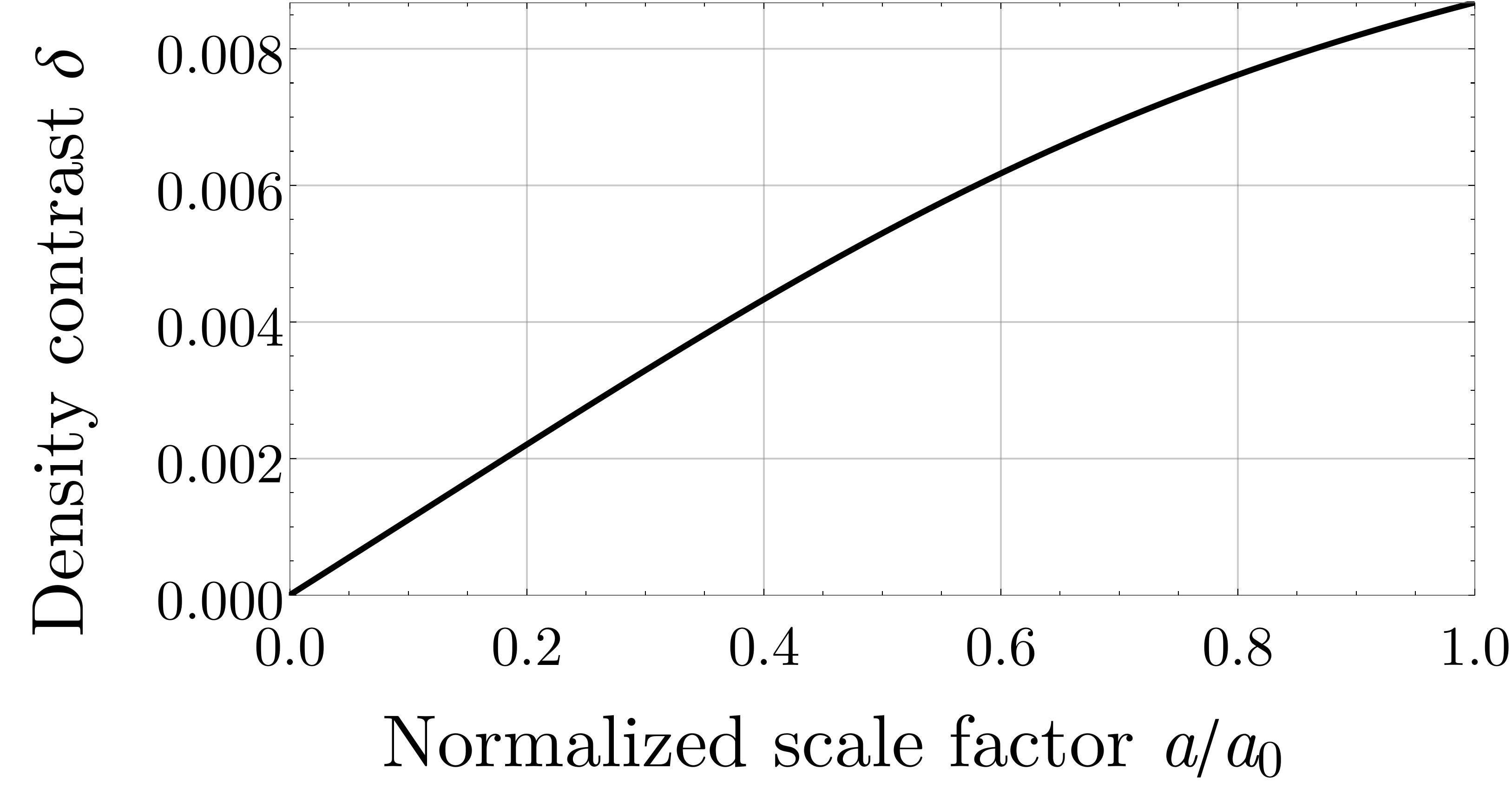}
	\caption{Density contrast for a homogeneous overdensity as a function of the normalized scale factor for a valid $\beta_1 \beta_2$ model. As an example, we have set $\delta=10^{-5}$ at $a/a_0 \simeq 10^{-3}$, approximately corresponding to the time of photon decoupling. Despite $\delta$ being small, note that the results hold nonlinearly.}
	\label{fig:delta-a}
\end{figure}

In this paper, we have been concerned with a perturbative solution of the full nonlinear bimetric equations of motion. To enable comparison with the results of linear perturbation theory, we define the density contrast,
\begin{equation}
	\delta := (\rho -\wb{\rho})/ \wb{\rho},
\end{equation}
between the background matter density $\wb{\rho}$ and the (overdense or underdense) region with density $\rho$. As an example, at redshift $z = z_* =1100$ (approximately corresponding to the time of photon decoupling), we set,
\begin{equation}
	\delta_* = 10^{-5}, \quad \left. \frac{d \log \delta}{d \log a} \right|_* = 1, \quad a_* = \wb{a}_*.
\end{equation}
The last equation states that the scale factors of the overdensity and the background are equal at $z_*$. The equation of motion for the background scale factor is \eqref{eq:BRFriedmann} with $K=0$ whereas for the overdense region $K=\delta K >0$. The density contrast grows linearly with the scale factor during matter domination. An example is plotted in figure \ref{fig:delta-a} for a valid $\beta_1 \beta_2$ model with $B_1 \simeq 13.1$, $\Omega_{\mathrm{M},0} = 0.30$, and $H_0 = 70 \, \mathrm{km/s/Mpc}$; starting with $\delta = 10^{-5}$ at $z=1100$, it grows to $\delta \simeq 8.5 \times 10^{-3}$ at $z=0$. 

\noindent The growth rate,
\begin{equation}
f_g := \frac{d \log \delta}{d \log a},
\end{equation}
is a commonly used parameter, quantifying the difference in structure growth between GR and modified theories of gravity \cite{Wang:1998gt,Linder:2005in,Huterer:2006mva,Ferreira:2010sz}. An approximation of $f_g$ is,
\begin{equation}
\label{eq:gamma_Def}
f_g \simeq \left[\left(\frac{H_0}{H}\right)^2 \Omega_\mathrm{M}\right]^\gamma,
\end{equation}
where the growth index $\gamma$ is the best fit parameter of the right-hand side to the left-hand side of \eqref{eq:gamma_Def}. 

\section{Graphical solution of bimetric cosmology}
\label{sec:graphical}
Here, we present a graphical method to determine the evolution of bimetric background cosmologies. With the inside of a homogeneous overdensity reducing to such a solution, the method applies equally well to the evolution of such perturbations.

\begin{figure}[tbp]
	\centering
	\includegraphics[width=\linewidth]{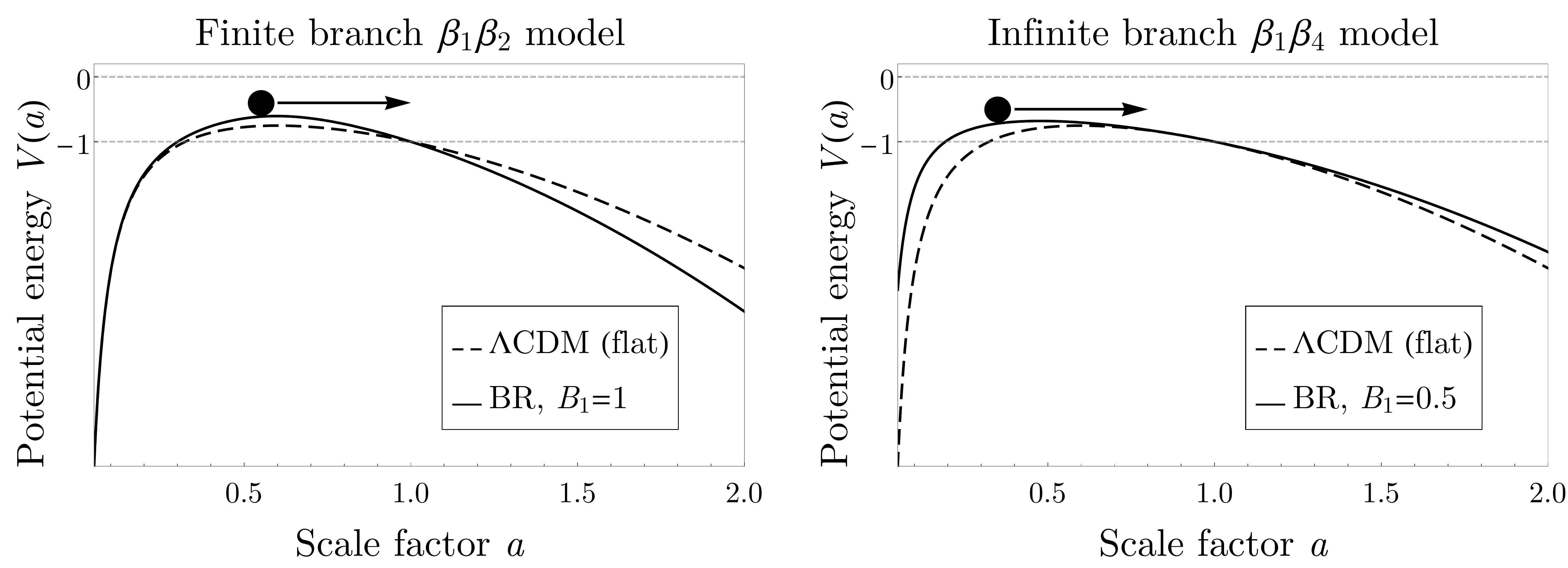}
	\caption{Potential profiles for two bimetric models and the flat ($K=0$) $\Lambda \mathrm{CDM}$ reference. The bimetric and GR potentials cross at $a=1$ (today) because we have imposed $H_0^{\mathrm{GR}} = H_0^{\mathrm{BR}}$.}
	\label{fig:graphical-sol}
	\end{figure}

Assuming that the metrics are homogeneous and isotropic in the same coordinates, the general Ansatz is \eqref{eq:BRFLRWAnsatz}. The Friedmann-like equation \eqref{eq:BRFriedmann} takes the form of an energy conservation law where $a$ is thought of as the coordinate of a particle moving in a potential $V$,
\begin{equation}
\label{eq:energycons}
T(\dot{a}) + V(a) = E,
\end{equation}
with the kinetic and potential energy defined by, respectively,
\begin{equation}
\label{eq:KEPE}
T(\dot{a})= \frac{\dot{a}^2}{H_0^2}, \quad V(a) = -a^2 \left[\Omega_{\mathrm{M}}(a) + \Omega_{\mathrm{DE}}(a)\right], \quad E = -\frac{K}{H_0^2}.
\end{equation}
An equation of the form \eqref{eq:energycons} can be solved graphically by imagining a particle sliding along the profile of the potential. To complete the method, $\Omega_{\mathrm{M}}$ and $\Omega_{\mathrm{DE}}$ (hence $y$) must be expressed as functions of $a$. From conservation of stress--energy and the equation of state $P=w\rho$, one derives,
\begin{equation}
\label{eq:rhocosmo}
\Omega_{\mathrm{M}} = \Omega_{\mathrm{M},0} \left(a/a_0\right)^{-3 (1+w)}.
\end{equation}
Equation \eqref{eq:ypoly}, can be solved for $y$ as a function of $a$ which can be reinserted in \eqref{eq:cosmodefs}. It is important to require that $y$ is real, possibly discarding some solution branches. Having expressed $\Omega_{\mathrm{M}}$ and $\Omega_{\mathrm{DE}}$ as functions of $a$, one can plot the potential $V(a)$ as a function of $a$ and the evolution of $a$ can be read off graphically as a particle sliding along the potential profile, see figure \ref{fig:graphical-sol}. 

If desired, one can impose constraints on the allowed regions of motion of $a$ in the potential. Requiring a principal square-root amounts to $y>0$. To avoid the Higuchi ghost, the requirement is that \cite{Fasiello:2013woa,Konnig:2015lfa},
\begin{equation}
\label{eq:Higuchi}
F := 3 B_3 y^4 + 2 \left(3 B_2 -B_4\right)y^3 + 3 \left(B_1 - B_3\right)y^2 + B_1 \geq 0.
\end{equation}
Since $y$ is known as a function of $a$, $F$ can be expressed as a function of $a$, possibly restricting the allowed values of $a$.

Note that,
\begin{equation}
\label{eq:X}
\wt{\alpha} = \dot{\wt{a}}/\dot{a} = y + a \, \partial y/\partial a.
\end{equation}
The right-hand side can be straightforwardly computed since $y$ is known as a function of $a$. Thus, from \eqref{eq:X} it follows that if the evolution of $a$ (i.e., $\dot{a}$) is known, then the evolution of $\wt{a}$ (i.e., $\dot{\wt{a}}$) can be computed immediately. For example, if $\dot{a}>0$ and $y + a \, \partial y/\partial a >0$, then $\wt{a}$ is expanding, $\dot{\wt{a}}>0$. Note that $\wt{\alpha}$ may hit zero at some $a$, at which point $\det S =0$. To exclude such solutions one can simply discard those solutions branches, or ensure appropriate initial conditions and $K$, or push this point sufficiently far away into the future.
Summarizing the method:\\

\noindent (1) Determine $\Omega_{\mathrm{M}}$ as a function of $a$ by solving the stress--energy conservation equation.\\

\noindent (2) Determine $y$ as a function of $a$ by solving the quartic polynomial \eqref{eq:ypoly}, selecting only the real solutions.\\

\noindent (3)  Reinsert the solution of $y$ in $\Omega_{\mathrm{DE}}$ \eqref{eq:cosmodefs}.\\

\noindent (4) Require a principal square root, that is, $y>0$. This may involve selecting different branches, depending on the value of the $\beta$-parameters. \\

\noindent (5) Plot $F$ as a function of $a$ \eqref{eq:Higuchi}. Requiring $F(a) \geq 0$ adds a new restriction.\\

\noindent (6) Plot the potential energy $V(a)$ as a function of $a$. This gives an exact graphical solution of the evolution of $a$, given some initial condition.\\

\noindent (7) Plot $\wt{\alpha}$ as a function of $a$ \eqref{eq:X}. This determines the evolution of $\wt{a}$, given the evolution of $a$. If $\wt{\alpha}$ hits zero at some $a$, the solution should be prevented from getting there. For example, pushing it into the future, or ensuring appropriate initial conditions and $K$, or ruling out the branch.\\

\noindent For details on a similar method, see \cite{Mortsell:2016too,Mortsell:2017fog}.

\section{Detailed calculations}
\label{sec:details}
Here, we supplement the missing steps in section \ref{sec:LTB_hom_lapse}.

\paragraph{Steps \eqref{eq:falpha}-\eqref{eq:LTBeqns}.} With \eqref{eq:falpha}, the $E^t{}_t$ equation with pressureless dust stress--energy \eqref{eq:Tgdust} can be written,
\begin{equation}
\label{eq:Ett}
\partial_r U = \kappa_g \rho r^2 B^2 \partial_r (rB) + \kappa_g \beta_1 \partial_r \left(r^3 B^2 \wt{B}\right)/\ell^2, \quad U:= r^3 B \left( \dot{B}^2 + k/r^2\right).
\end{equation}
For simplicity, only the bimetric $\beta_1$-terms are written out. The full result is shown in \eqref{eq:Friedmann_g}. Integrating,
\begin{equation}
\label{eq:Usol1}
U = \kappa_g M + \kappa_g \beta_1 r^3 B^2 \wt{B}/ \ell^2 +b_1(t), \quad M := \int_{0}^r dr \rho r^2 B^2 \partial_r (rB),
\end{equation}
with $b_1(t)$ a generic function of $t$. Due to conservation of stress--energy \eqref{eq:stressenCons},
\begin{equation}
\partial_t M =0 \quad \Rightarrow \quad M = M(r).
\end{equation}
Using \eqref{eq:falpha}, the $E^r{}_r$ equation is,
\begin{equation}
\partial_t U  = \kappa_g \beta_1 r^3 \partial_t (B^2 \wt{B}) / \ell^2,
\end{equation}
and integrating,
\begin{equation}
\label{eq:Usol2}
U = \kappa_g \beta_1 r^3 B^2 \wt{B} / \ell^2 + \kappa_g b_2(r).
\end{equation}
Combining \eqref{eq:Usol1} and \eqref{eq:Usol2}, $b_1 = 0$ and $b_2 = M$, and reinserting all the $\beta$-parameters,
\begin{equation}
\label{eq:Friedmann_g}
H^2 + \frac{k}{r^2 B^2} = \kappa_g \frac{M}{r^3 B^3} + \frac{\kappa_g}{\ell^2} \left(\frac{\beta_0}{3} + \beta_1 y + \beta_2 y^2 + \frac{\beta_3}{3}y^3\right), \quad H:= \frac{\dot{B}}{B}, \quad y := \frac{\wt{B}}{B}.
\end{equation}
Note the similarity between \eqref{eq:Friedmann_g} and the Friedmann equation of bimetric FLRW cosmology. Defining,
\begin{equation}
\wt{U} := r^3 \wt{B} \left(\dot{B}^2 + k/r^2 \right),
\end{equation}
we repeat the steps \eqref{eq:Ett}-\eqref{eq:Friedmann_g} in the $f$-sector and obtain,
\begin{equation}
\label{eq:Friedmann_f}
H^2 + \frac{k}{r^2 B^2} = \frac{\kappa_f}{\ell^2} \left(\frac{\beta_1}{3} y^{-1} + \beta_2 + \beta_3 y + \frac{\beta_4}{3} y^2 \right).
\end{equation}
Subtracting \eqref{eq:Friedmann_g} from \eqref{eq:Friedmann_f} yields a quartic polynomial equation in $y$,
\begin{equation}
\label{eq:y_poly}
\frac{1}{\ell^2} \left[-\frac{\beta_3}{3} y^3 + \left(\frac{\beta_4}{3\kappa} - \beta_2\right)y^2+ \left(\frac{\beta_3}{\kappa} - \beta_1 \right)y + \left(\frac{\beta_2}{\kappa} - \frac{\beta_0}{3}\right) + \frac{\beta_1}{3\kappa} y^{-1}\right]= \frac{M}{r^3 B^3},
\end{equation}
with the definition $\kappa := \kappa_g / \kappa_f$. 

\paragraph{Steps \eqref{eq:B_separable}-\eqref{eq:bimetricFLRW}.} Plugging \eqref{eq:B_separable} into \eqref{eq:flapse} gives,
\begin{equation}
\wt{B}(t,r) = \wt{a}(t)b(r) + c(r). 
\end{equation}
Writing out the metrics,
\begin{subequations}
	\begin{align}
	ds_g^2 &= - d t^2 + \frac{a^2 \left(r b\right)' \hspace{0.1mm}^2}{1-k}d r^2 + a^2 b^2 r^2 d \Omega^2,\\
	ds_f^2  &= - \frac{\dot{\wt{a}}^2}{\dot{a}^2} d t^2 + \frac{\left[r (\wt{a}b + c)\right]' \hspace{0.1mm}^2}{1-k} d r^2 + \left(\wt{a}b + c\right)^2 r^2 d \Omega^2.
	\end{align}
\end{subequations}
Transforming coordinates $r \to rb(r)$,
\begin{subequations}
	\begin{align}
	ds_g^2 &= - d t^2 + a(t)^2 \left(\frac{d r^2}{1-k(r)} + r^2 d \Omega^2\right),\\
	ds_f^2 &= - \frac{\dot{\wt{a}}^2(t)}{\dot{a}^2(t)} d t^2 + \frac{\left[\wt{a}(t)+c(r)'\right]^2}{1-k(r)} d r^2 + \left[\wt{a}(t) + \frac{c(r)}{r}\right]^2 r^2 d \Omega^2.
	\end{align}
\end{subequations}
Calculating $\partial_r E^r{}_r$ and setting to zero gives a first-order differential equation in $r$. Solving for $k(r)$ yields a time-dependent expression containing $a(t)$ and $\wt{a}(t)$. Only if $c(r)=0$, the $t$-dependence goes away and,
\begin{equation}
k(r) = K r^2 , \quad c(r)=0,
\end{equation}
with constant $K$. Hence, the bimetric FLRW Ansatz is finally recovered,
\begin{subequations}
	\begin{alignat}{2}
	ds_g^2 &= - d t^2 + a(t)^2 \left(\frac{d r^2}{1-K r^2} + r^2 d \Omega^2\right),& \quad K &= \mathrm{const.},\\
	ds_f^2 &= - \wt{\alpha}^2(t)d t^2 + \wt{a}(t)^2 \left(\frac{d r^2}{1-K r^2} + r^2 d \Omega^2\right),& \quad \wt{\alpha}(t) &= \frac{\dot{\wt{a}}(t)}{\dot{a}(t)} .
	\end{alignat}
\end{subequations}

\bibliographystyle{JHEP}
\bibliography{biblio}

\providecommand{\href}[2]{#2}\begingroup\raggedright\begin{thebibliography}{10}

\bibitem{Hassan:2011zd}
S.~F. Hassan and R.~A. Rosen, \emph{{Bimetric Gravity from Ghost-free Massive
  Gravity}}, \href{http://dx.doi.org/10.1007/JHEP02(2012)126}{\emph{JHEP} {\bf
  02} (2012) 126}, [\href{http://arxiv.org/abs/1109.3515}{{\tt 1109.3515}}].

\bibitem{Hassan:2011ea}
S.~F. Hassan and R.~A. Rosen, \emph{{Confirmation of the Secondary Constraint
  and Absence of Ghost in Massive Gravity and Bimetric Gravity}},
  \href{http://dx.doi.org/10.1007/JHEP04(2012)123}{\emph{JHEP} {\bf 04} (2012)
  123}, [\href{http://arxiv.org/abs/1111.2070}{{\tt 1111.2070}}].

\bibitem{Hassan:2012wr}
S.~F. Hassan, A.~Schmidt-May and M.~von Strauss, \emph{{On Consistent Theories
  of Massive Spin-2 Fields Coupled to Gravity}},
  \href{http://dx.doi.org/10.1007/JHEP05(2013)086}{\emph{JHEP} {\bf 05} (2013)
  086}, [\href{http://arxiv.org/abs/1208.1515}{{\tt 1208.1515}}].

\bibitem{Hassan:2018mbl}
S.~F. Hassan and A.~Lundkvist, \emph{{Analysis of constraints and their algebra
  in bimetric theory}},
  \href{http://dx.doi.org/10.1007/JHEP08(2018)182}{\emph{JHEP} {\bf 08} (2018)
  182}, [\href{http://arxiv.org/abs/1802.07267}{{\tt 1802.07267}}].

\bibitem{deRham:2010ik}
C.~de~Rham and G.~Gabadadze, \emph{{Generalization of the Fierz-Pauli Action}},
  \href{http://dx.doi.org/10.1103/PhysRevD.82.044020}{\emph{Phys. Rev.} {\bf
  D82} (2010) 044020}, [\href{http://arxiv.org/abs/1007.0443}{{\tt
  1007.0443}}].

\bibitem{deRham:2010kj}
C.~de~Rham, G.~Gabadadze and A.~J. Tolley, \emph{{Resummation of Massive
  Gravity}},
  \href{http://dx.doi.org/10.1103/PhysRevLett.106.231101}{\emph{Phys. Rev.
  Lett.} {\bf 106} (2011) 231101}, [\href{http://arxiv.org/abs/1011.1232}{{\tt
  1011.1232}}].

\bibitem{Hassan:2011hr}
S.~F. Hassan and R.~A. Rosen, \emph{{Resolving the Ghost Problem in non-Linear
  Massive Gravity}},
  \href{http://dx.doi.org/10.1103/PhysRevLett.108.041101}{\emph{Phys. Rev.
  Lett.} {\bf 108} (2012) 041101}, [\href{http://arxiv.org/abs/1106.3344}{{\tt
  1106.3344}}].

\bibitem{Hassan:2014vja}
S.~F. Hassan, A.~Schmidt-May and M.~von Strauss, \emph{{Particular Solutions in
  Bimetric Theory and Their Implications}},
  \href{http://dx.doi.org/10.1142/S0218271814430020}{\emph{Int. J. Mod. Phys.}
  {\bf D23} (2014) 1443002}, [\href{http://arxiv.org/abs/1407.2772}{{\tt
  1407.2772}}].

\bibitem{Fierz:1939ix}
M.~Fierz and W.~Pauli, \emph{{On relativistic wave equations for particles of
  arbitrary spin in an electromagnetic field}},
  \href{http://dx.doi.org/10.1098/rspa.1939.0140}{\emph{Proc. Roy. Soc. Lond.}
  {\bf A173} (1939) 211--232}.

\bibitem{deRham:2014zqa}
C.~de~Rham, \emph{{Massive Gravity}},
  \href{http://dx.doi.org/10.12942/lrr-2014-7}{\emph{Living Rev. Rel.} {\bf 17}
  (2014) 7}, [\href{http://arxiv.org/abs/1401.4173}{{\tt 1401.4173}}].

\bibitem{Schmidt-May:2015vnx}
A.~Schmidt-May and M.~von Strauss, \emph{{Recent developments in bimetric
  theory}}, \href{http://dx.doi.org/10.1088/1751-8113/49/18/183001}{\emph{J.
  Phys.} {\bf A49} (2016) 183001}, [\href{http://arxiv.org/abs/1512.00021}{{\tt
  1512.00021}}].

\bibitem{vonStrauss:2011mq}
M.~von Strauss, A.~Schmidt-May, J.~Enander, E.~Mortsell and S.~F. Hassan,
  \emph{{Cosmological Solutions in Bimetric Gravity and their Observational
  Tests}}, \href{http://dx.doi.org/10.1088/1475-7516/2012/03/042}{\emph{JCAP}
  {\bf 1203} (2012) 042}, [\href{http://arxiv.org/abs/1111.1655}{{\tt
  1111.1655}}].

\bibitem{Volkov:2011an}
M.~S. Volkov, \emph{{Cosmological solutions with massive gravitons in the
  bigravity theory}},
  \href{http://dx.doi.org/10.1007/JHEP01(2012)035}{\emph{JHEP} {\bf 01} (2012)
  035}, [\href{http://arxiv.org/abs/1110.6153}{{\tt 1110.6153}}].

\bibitem{Comelli:2011zm}
D.~Comelli, M.~Crisostomi, F.~Nesti and L.~Pilo, \emph{{FRW Cosmology in Ghost
  Free Massive Gravity}}, \href{http://dx.doi.org/10.1007/JHEP06(2012)020,
  10.1007/JHEP03(2012)067}{\emph{JHEP} {\bf 03} (2012) 067},
  [\href{http://arxiv.org/abs/1111.1983}{{\tt 1111.1983}}].

\bibitem{Akrami:2012vf}
Y.~Akrami, T.~S. Koivisto and M.~Sandstad, \emph{{Accelerated expansion from
  ghost-free bigravity: a statistical analysis with improved generality}},
  \href{http://dx.doi.org/10.1007/JHEP03(2013)099}{\emph{JHEP} {\bf 03} (2013)
  099}, [\href{http://arxiv.org/abs/1209.0457}{{\tt 1209.0457}}].

\bibitem{Akrami:2013pna}
Y.~Akrami, T.~S. Koivisto and M.~Sandstad, \emph{{Cosmological constraints on
  ghost-free bigravity: background dynamics and late-time acceleration}},  in
  \emph{{Proceedings, 13th Marcel Grossmann Meeting on Recent Developments in
  Theoretical and Experimental General Relativity, Astrophysics, and
  Relativistic Field Theories (MG13): Stockholm, Sweden, July 1-7, 2012}},
  pp.~1252--1254, 2015.
\newblock \href{http://arxiv.org/abs/1302.5268}{{\tt 1302.5268}}.
\newblock \href{http://dx.doi.org/10.1142/9789814623995_0138}{DOI}.

\bibitem{Nersisyan:2015oha}
H.~Nersisyan, Y.~Akrami and L.~Amendola, \emph{{Consistent metric combinations
  in cosmology of massive bigravity}},
  \href{http://dx.doi.org/10.1103/PhysRevD.92.104034}{\emph{Phys. Rev.} {\bf
  D92} (2015) 104034}, [\href{http://arxiv.org/abs/1502.03988}{{\tt
  1502.03988}}].

\bibitem{Enander:2015kda}
J.~Enander and E.~Mortsell, \emph{{On stars, galaxies and black holes in
  massive bigravity}},
  \href{http://dx.doi.org/10.1088/1475-7516/2015/11/023}{\emph{JCAP} {\bf 1511}
  (2015) 023}, [\href{http://arxiv.org/abs/1507.00912}{{\tt 1507.00912}}].

\bibitem{Aoki:2016zgp}
K.~Aoki and S.~Mukohyama, \emph{{Massive gravitons as dark matter and
  gravitational waves}},
  \href{http://dx.doi.org/10.1103/PhysRevD.94.024001}{\emph{Phys. Rev.} {\bf
  D94} (2016) 024001}, [\href{http://arxiv.org/abs/1604.06704}{{\tt
  1604.06704}}].

\bibitem{Babichev:2016hir}
E.~Babichev, L.~Marzola, M.~Raidal, A.~Schmidt-May, F.~Urban, H.~Veermäe
  et~al., \emph{{Bigravitational origin of dark matter}},
  \href{http://dx.doi.org/10.1103/PhysRevD.94.084055}{\emph{Phys. Rev.} {\bf
  D94} (2016) 084055}, [\href{http://arxiv.org/abs/1604.08564}{{\tt
  1604.08564}}].

\bibitem{Babichev:2016bxi}
E.~Babichev, L.~Marzola, M.~Raidal, A.~Schmidt-May, F.~Urban, H.~Veermäe
  et~al., \emph{{Heavy spin-2 Dark Matter}},
  \href{http://dx.doi.org/10.1088/1475-7516/2016/09/016}{\emph{JCAP} {\bf 1609}
  (2016) 016}, [\href{http://arxiv.org/abs/1607.03497}{{\tt 1607.03497}}].

\bibitem{Comelli:2012db}
D.~Comelli, M.~Crisostomi and L.~Pilo, \emph{{Perturbations in Massive Gravity
  Cosmology}}, \href{http://dx.doi.org/10.1007/JHEP06(2012)085}{\emph{JHEP}
  {\bf 06} (2012) 085}, [\href{http://arxiv.org/abs/1202.1986}{{\tt
  1202.1986}}].

\bibitem{Khosravi:2012rk}
N.~Khosravi, H.~R. Sepangi and S.~Shahidi, \emph{{Massive cosmological scalar
  perturbations}},
  \href{http://dx.doi.org/10.1103/PhysRevD.86.043517}{\emph{Phys. Rev.} {\bf
  D86} (2012) 043517}, [\href{http://arxiv.org/abs/1202.2767}{{\tt
  1202.2767}}].

\bibitem{Berg:2012kn}
M.~Berg, I.~Buchberger, J.~Enander, E.~Mortsell and S.~Sjors, \emph{{Growth
  Histories in Bimetric Massive Gravity}},
  \href{http://dx.doi.org/10.1088/1475-7516/2012/12/021}{\emph{JCAP} {\bf 1212}
  (2012) 021}, [\href{http://arxiv.org/abs/1206.3496}{{\tt 1206.3496}}].

\bibitem{Fasiello:2013woa}
M.~Fasiello and A.~J. Tolley, \emph{{Cosmological Stability Bound in Massive
  Gravity and Bigravity}},
  \href{http://dx.doi.org/10.1088/1475-7516/2013/12/002}{\emph{JCAP} {\bf 1312}
  (2013) 002}, [\href{http://arxiv.org/abs/1308.1647}{{\tt 1308.1647}}].

\bibitem{Konnig:2014dna}
F.~Könnig and L.~Amendola, \emph{{Instability in a minimal bimetric gravity
  model}}, \href{http://dx.doi.org/10.1103/PhysRevD.90.044030}{\emph{Phys.
  Rev.} {\bf D90} (2014) 044030}, [\href{http://arxiv.org/abs/1402.1988}{{\tt
  1402.1988}}].

\bibitem{Comelli:2014bqa}
D.~Comelli, M.~Crisostomi and L.~Pilo, \emph{{FRW Cosmological Perturbations in
  Massive Bigravity}},
  \href{http://dx.doi.org/10.1103/PhysRevD.90.084003}{\emph{Phys. Rev.} {\bf
  D90} (2014) 084003}, [\href{http://arxiv.org/abs/1403.5679}{{\tt
  1403.5679}}].

\bibitem{DeFelice:2014nja}
A.~De~Felice, A.~E. Gümrükçüoğlu, S.~Mukohyama, N.~Tanahashi and
  T.~Tanaka, \emph{{Viable cosmology in bimetric theory}},
  \href{http://dx.doi.org/10.1088/1475-7516/2014/06/037}{\emph{JCAP} {\bf 1406}
  (2014) 037}, [\href{http://arxiv.org/abs/1404.0008}{{\tt 1404.0008}}].

\bibitem{Solomon:2014dua}
A.~R. Solomon, Y.~Akrami and T.~S. Koivisto, \emph{{Linear growth of structure
  in massive bigravity}},
  \href{http://dx.doi.org/10.1088/1475-7516/2014/10/066}{\emph{JCAP} {\bf 1410}
  (2014) 066}, [\href{http://arxiv.org/abs/1404.4061}{{\tt 1404.4061}}].

\bibitem{Konnig:2014xva}
F.~Koennig, Y.~Akrami, L.~Amendola, M.~Motta and A.~R. Solomon, \emph{{Stable
  and unstable cosmological models in bimetric massive gravity}},
  \href{http://dx.doi.org/10.1103/PhysRevD.90.124014}{\emph{Phys. Rev.} {\bf
  D90} (2014) 124014}, [\href{http://arxiv.org/abs/1407.4331}{{\tt
  1407.4331}}].

\bibitem{Lagos:2014lca}
M.~Lagos and P.~G. Ferreira, \emph{{Cosmological perturbations in massive
  bigravity}},
  \href{http://dx.doi.org/10.1088/1475-7516/2014/12/026}{\emph{JCAP} {\bf 1412}
  (2014) 026}, [\href{http://arxiv.org/abs/1410.0207}{{\tt 1410.0207}}].

\bibitem{Enander:2015vja}
J.~Enander, Y.~Akrami, E.~Mörtsell, M.~Renneby and A.~R. Solomon,
  \emph{{Integrated Sachs-Wolfe effect in massive bigravity}},
  \href{http://dx.doi.org/10.1103/PhysRevD.91.084046}{\emph{Phys. Rev.} {\bf
  D91} (2015) 084046}, [\href{http://arxiv.org/abs/1501.02140}{{\tt
  1501.02140}}].

\bibitem{Konnig:2015lfa}
F.~Könnig, \emph{{Higuchi Ghosts and Gradient Instabilities in Bimetric
  Gravity}}, \href{http://dx.doi.org/10.1103/PhysRevD.91.104019}{\emph{Phys.
  Rev.} {\bf D91} (2015) 104019}, [\href{http://arxiv.org/abs/1503.07436}{{\tt
  1503.07436}}].

\bibitem{Kenna-Allison:2018izo}
M.~Kenna-Allison, A.~E. Gümrükçüoǧlu and K.~Koyama, \emph{{Viability of
  bigravity cosmology}},
  \href{http://dx.doi.org/10.1103/PhysRevD.99.104032}{\emph{Phys. Rev.} {\bf
  D99} (2019) 104032}, [\href{http://arxiv.org/abs/1812.05496}{{\tt
  1812.05496}}].

\bibitem{Akrami:2015qga}
Y.~Akrami, S.~F. Hassan, F.~Könnig, A.~Schmidt-May and A.~R. Solomon,
  \emph{{Bimetric gravity is cosmologically viable}},
  \href{http://dx.doi.org/10.1016/j.physletb.2015.06.062}{\emph{Phys. Lett.}
  {\bf B748} (2015) 37--44}, [\href{http://arxiv.org/abs/1503.07521}{{\tt
  1503.07521}}].

\bibitem{Vainshtein:1972sx}
A.~I. Vainshtein, \emph{{To the problem of nonvanishing gravitation mass}},
  \href{http://dx.doi.org/10.1016/0370-2693(72)90147-5}{\emph{Phys. Lett.} {\bf
  39B} (1972) }.

\bibitem{Aoki:2015xqa}
K.~Aoki, K.-i. Maeda and R.~Namba, \emph{{Stability of the Early Universe in
  Bigravity Theory}},
  \href{http://dx.doi.org/10.1103/PhysRevD.92.044054}{\emph{Phys. Rev.} {\bf
  D92} (2015) 044054}, [\href{http://arxiv.org/abs/1506.04543}{{\tt
  1506.04543}}].

\bibitem{Mortsell:2015exa}
E.~Mortsell and J.~Enander, \emph{{Scalar instabilities in bimetric gravity:
  The Vainshtein mechanism and structure formation}},
  \href{http://dx.doi.org/10.1088/1475-7516/2015/10/044}{\emph{JCAP} {\bf 1510}
  (2015) 044}, [\href{http://arxiv.org/abs/1506.04977}{{\tt 1506.04977}}].

\bibitem{Luben:2019yyx}
M.~Lüben, A.~Schmidt-May and J.~Smirnov, \emph{{Vainshtein Screening in
  Bimetric Cosmology}},  \href{http://arxiv.org/abs/1912.09449}{{\tt
  1912.09449}}.

\bibitem{Weinberg:2008zzc}
S.~Weinberg, \emph{{Cosmology}}.
\newblock Oxford, UK: Oxford Univ. Pr. (2008) 593 p, 2008.

\bibitem{Ijjas:2018cdm}
A.~Ijjas, F.~Pretorius and P.~J. Steinhardt, \emph{{Stability and the Gauge
  Problem in Non-Perturbative Cosmology}},
  \href{http://dx.doi.org/10.1088/1475-7516/2019/01/015}{\emph{JCAP} {\bf 1901}
  (2019) 015}, [\href{http://arxiv.org/abs/1809.07010}{{\tt 1809.07010}}].

\bibitem{Kocic:2018ddp}
M.~Kocic, \emph{{Geometric mean of bimetric spacetimes}},
  \href{http://arxiv.org/abs/1803.09752}{{\tt 1803.09752}}.

\bibitem{Kocic:2018yvr}
M.~Kocic, \emph{{Causal propagation of constraints in bimetric relativity in
  standard 3+1 form}},
  \href{http://dx.doi.org/10.1007/JHEP10(2019)219}{\emph{JHEP} {\bf 10} (2019)
  219}, [\href{http://arxiv.org/abs/1804.03659}{{\tt 1804.03659}}].

\bibitem{cbssn}
F.~Torsello, M.~Kocic, M.~Högås and E.~Mörtsell, \emph{{Covariant BSSN
  formulation in bimetric relativity}},
  \href{http://dx.doi.org/10.1088/1361-6382/ab56fc}{\emph{Class. Quant. Grav.}
  {\bf 37} (2020) 025013}, [\href{http://arxiv.org/abs/1904.07869}{{\tt
  1904.07869}}].

\bibitem{Kocic:2019zdy}
M.~Kocic, A.~Lundkvist and F.~Torsello, \emph{{On the ratio of lapses in
  bimetric relativity}},
  \href{http://dx.doi.org/10.1088/1361-6382/ab497a}{\emph{Class. Quant. Grav.}
  {\bf 36} (2019) 225013}, [\href{http://arxiv.org/abs/1903.09646}{{\tt
  1903.09646}}].

\bibitem{polytropes}
M.~Kocic, F.~Torsello, M.~Högås and E.~Mortsell, \emph{{Spherical dust
  collapse in bimetric relativity: Bimetric polytropes}},
  \href{http://arxiv.org/abs/1904.08617}{{\tt 1904.08617}}.

\bibitem{meangauge}
F.~Torsello, \emph{{The mean gauges in bimetric relativity}},
  \href{http://dx.doi.org/10.1088/1361-6382/ab4ccf}{\emph{Class. Quant. Grav.}
  {\bf 36} (2019) 235010}, [\href{http://arxiv.org/abs/1904.09297}{{\tt
  1904.09297}}].

\bibitem{bimEX}
F.~Torsello, \emph{{$\mathtt{bimEX}$: A Mathematica package for exact
  computations in $3+1$ bimetric relativity}},
  \href{http://dx.doi.org/10.1016/j.cpc.2019.106948}{\emph{Comput. Phys.
  Commun.} {\bf 247} (2020) 106948},
  [\href{http://arxiv.org/abs/1904.10464}{{\tt 1904.10464}}].

\bibitem{Laureijs:2011gra}
{\scshape EUCLID} collaboration, R.~Laureijs et~al., \emph{{Euclid Definition
  Study Report}},  \href{http://arxiv.org/abs/1110.3193}{{\tt 1110.3193}}.

\bibitem{Amendola:2012ys}
{\scshape Euclid Theory Working Group} collaboration, L.~Amendola et~al.,
  \emph{{Cosmology and fundamental physics with the Euclid satellite}},
  \href{http://dx.doi.org/10.12942/lrr-2013-6}{\emph{Living Rev. Rel.} {\bf 16}
  (2013) 6}, [\href{http://arxiv.org/abs/1206.1225}{{\tt 1206.1225}}].

\bibitem{Amendola:2016saw}
L.~Amendola et~al., \emph{{Cosmology and fundamental physics with the Euclid
  satellite}}, \href{http://dx.doi.org/10.1007/s41114-017-0010-3}{\emph{Living
  Rev. Rel.} {\bf 21} (2018) 2}, [\href{http://arxiv.org/abs/1606.00180}{{\tt
  1606.00180}}].

\bibitem{Dhawan:2017leu}
S.~Dhawan, A.~Goobar, E.~Mörtsell, R.~Amanullah and U.~Feindt,
  \emph{{Narrowing down the possible explanations of cosmic acceleration with
  geometric probes}},
  \href{http://dx.doi.org/10.1088/1475-7516/2017/07/040}{\emph{JCAP} {\bf 1707}
  (2017) 040}, [\href{http://arxiv.org/abs/1705.05768}{{\tt 1705.05768}}].

\bibitem{Mortsell:2018mfj}
E.~Mörtsell and S.~Dhawan, \emph{{Does the Hubble constant tension call for
  new physics?}},
  \href{http://dx.doi.org/10.1088/1475-7516/2018/09/025}{\emph{JCAP} {\bf 1809}
  (2018) 025}, [\href{http://arxiv.org/abs/1801.07260}{{\tt 1801.07260}}].

\bibitem{Boulware:1973my}
D.~G. Boulware and S.~Deser, \emph{{Can gravitation have a finite range?}},
  \href{http://dx.doi.org/10.1103/PhysRevD.6.3368}{\emph{Phys. Rev.} {\bf D6}
  (1972) 3368--3382}.

\bibitem{Hassan:2017ugh}
S.~F. Hassan and M.~Kocic, \emph{{On the local structure of spacetime in
  ghost-free bimetric theory and massive gravity}},
  \href{http://dx.doi.org/10.1007/JHEP05(2018)099}{\emph{JHEP} {\bf 05} (2018)
  099}, [\href{http://arxiv.org/abs/1706.07806}{{\tt 1706.07806}}].

\bibitem{Lemaitre:1933}
G.~Lemaître, \emph{{L'Univers en expansion}}, {\emph{Annales de la Societe
  Scietique de Bruxelles} {\bf 53} (1933) }.

\bibitem{Tolman:1934}
R.~C. Tolman, \emph{{Effect of Inhomogeneity on Cosmological Models}},
  {\emph{Proceedings of the National Academy of Science} {\bf 20} (1934)
  169--176}.

\bibitem{Bondi:1947fta}
H.~Bondi, \emph{{Spherically symmetrical models in general relativity}},
  \href{http://dx.doi.org/10.1093/mnras/107.5-6.410}{\emph{Mon. Not. Roy.
  Astron. Soc.} {\bf 107} (1947) 410--425}.

\bibitem{Oppenheimer:1939ue}
J.~R. Oppenheimer and H.~Snyder, \emph{{On Continued gravitational
  contraction}}, \href{http://dx.doi.org/10.1103/PhysRev.56.455}{\emph{Phys.
  Rev.} {\bf 56} (1939) 455--459}.

\bibitem{Jebsen:1921}
J.~T. Jebsen, \emph{{Über die allgemeinen kugelsymmetrischen Lösungen der
  Einsteinschen Gravitationsgleichungen im Vakuum}}, {\emph{Ark. Mat. Astr.
  Fys.} {\bf 15} (1921) }.

\bibitem{Jebsen2005}
J.~T. Jebsen, \emph{On the general spherically symmetric solutions of
  einstein's gravitational equations in vacuo},
  \href{http://dx.doi.org/10.1007/s10714-005-0168-y}{\emph{General Relativity
  and Gravitation} {\bf 37} (Dec, 2005) 2253--2259}.

\bibitem{Birkhoff:1923}
G.~Birkhoff and R.~Langer, \emph{{Relativity and Modern Physics}}.
\newblock Harvard University Press, 1923.

\bibitem{Kocic:2017hve}
M.~Kocic, M.~Högås, F.~Torsello and E.~Mortsell, \emph{{On Birkhoff's theorem
  in ghost-free bimetric theory}},  \href{http://arxiv.org/abs/1708.07833}{{\tt
  1708.07833}}.

\bibitem{Hogas:2019cpg}
M.~Högås, M.~Kocic, F.~Torsello and E.~Mörtsell, \emph{{Exact solutions for
  gravitational collapse in bimetric gravity}},
  \href{http://arxiv.org/abs/1905.09832}{{\tt 1905.09832}}.

\bibitem{Torsello:2017ouh}
F.~Torsello, M.~Kocic, M.~Högås and E.~Mortsell, \emph{{Spacetime symmetries
  and topology in bimetric relativity}},
  \href{http://dx.doi.org/10.1103/PhysRevD.97.084022}{\emph{Phys. Rev.} {\bf
  D97} (2018) 084022}, [\href{http://arxiv.org/abs/1710.06434}{{\tt
  1710.06434}}].

\bibitem{higham2008functions}
N.~Higham, \emph{Functions of Matrices: Theory and Computation}.
\newblock Society for Industrial and Applied Mathematics (SIAM, 3600 Market
  Street, Floor 6, Philadelphia, PA 19104), 2008.

\bibitem{Ruiz:2007rs}
M.~Ruiz, M.~Alcubierre and D.~Nunez, \emph{{Regularization of spherical and
  axisymmetric evolution codes in numerical relativity}},
  \href{http://dx.doi.org/10.1007/s10714-007-0522-3}{\emph{Gen. Rel. Grav.}
  {\bf 40} (2008) 159--182}, [\href{http://arxiv.org/abs/0706.0923}{{\tt
  0706.0923}}].

\bibitem{Smoot:1992td}
{\scshape COBE} collaboration, G.~F. Smoot et~al., \emph{{Structure in the COBE
  differential microwave radiometer first year maps}},
  \href{http://dx.doi.org/10.1086/186504}{\emph{Astrophys. J.} {\bf 396} (1992)
  L1--L5}.

\bibitem{Konnig:2013gxa}
F.~Koennig, A.~Patil and L.~Amendola, \emph{{Viable cosmological solutions in
  massive bimetric gravity}},
  \href{http://dx.doi.org/10.1088/1475-7516/2014/03/029}{\emph{JCAP} {\bf 1403}
  (2014) 029}, [\href{http://arxiv.org/abs/1312.3208}{{\tt 1312.3208}}].

\bibitem{Wang:1998gt}
L.-M. Wang and P.~J. Steinhardt, \emph{{Cluster abundance constraints on
  quintessence models}},
  \href{http://dx.doi.org/10.1086/306436}{\emph{Astrophys. J.} {\bf 508} (1998)
  483--490}, [\href{http://arxiv.org/abs/astro-ph/9804015}{{\tt
  astro-ph/9804015}}].

\bibitem{Linder:2005in}
E.~V. Linder, \emph{{Cosmic growth history and expansion history}},
  \href{http://dx.doi.org/10.1103/PhysRevD.72.043529}{\emph{Phys. Rev.} {\bf
  D72} (2005) 043529}, [\href{http://arxiv.org/abs/astro-ph/0507263}{{\tt
  astro-ph/0507263}}].

\bibitem{Huterer:2006mva}
D.~Huterer and E.~V. Linder, \emph{{Separating Dark Physics from Physical
  Darkness: Minimalist Modified Gravity vs. Dark Energy}},
  \href{http://dx.doi.org/10.1103/PhysRevD.75.023519}{\emph{Phys. Rev.} {\bf
  D75} (2007) 023519}, [\href{http://arxiv.org/abs/astro-ph/0608681}{{\tt
  astro-ph/0608681}}].

\bibitem{Ferreira:2010sz}
P.~G. Ferreira and C.~Skordis, \emph{{The linear growth rate of structure in
  Parametrized Post Friedmannian Universes}},
  \href{http://dx.doi.org/10.1103/PhysRevD.81.104020}{\emph{Phys. Rev.} {\bf
  D81} (2010) 104020}, [\href{http://arxiv.org/abs/1003.4231}{{\tt
  1003.4231}}].

\bibitem{Mortsell:2016too}
E.~Mortsell, \emph{{Cosmological histories from the Friedmann equation: The
  universe as a particle}},
  \href{http://dx.doi.org/10.1088/0143-0807/37/5/055603}{\emph{Eur. J. Phys.}
  {\bf 37} (2016) 055603}, [\href{http://arxiv.org/abs/1606.09556}{{\tt
  1606.09556}}].

\bibitem{Mortsell:2017fog}
E.~Mortsell, \emph{{Cosmological histories in bimetric gravity: A graphical
  approach}},
  \href{http://dx.doi.org/10.1088/1475-7516/2017/02/051}{\emph{JCAP} {\bf 1702}
  (2017) 051}, [\href{http://arxiv.org/abs/1701.00710}{{\tt 1701.00710}}].

\end{thebibliography}\endgroup

\end{document}